%% file: ms_revised_astroph.tex
\documentclass[iop,revtex4]{emulateapj}
\pdfoutput=1
\usepackage{natbib, fancyhdr,graphicx,amsmath}
\newcommand{\um}{${\rm \mu m}$~}

\newcommand{\ES}{$\Sigma_E$~}
\newcommand{\WS}{\textit{WISE}}
\def\hip{\textit{Hipparcos}}
\def\akari{\textit{AKARI}}
\def\iras{\textit{IRAS}}
\def\iso{\textit{ISO}}
\def\spitzer{\textit{Spitzer}}
\def\mass{\textit{2MASS}}
\shorttitle{\WS\ Debris Disks}
\shortauthors{Patel, Metchev, \& Heinze}

\begin{document}

\title{A Sensitive Identification of Warm Debris Disks in the Solar Neighborhood \\through Precise Calibration of Saturated \WS\ Photometry}


\author{Rahul I. Patel\altaffilmark{1}, Stanimir A. Metchev\altaffilmark{1,2}, Aren Heinze\altaffilmark{1}}
\altaffiltext{1}{Department of Physics \& Astronomy, Stony Brook University, 100 Nicolls Road, Stony Brook, New York 11794--3800}
\altaffiltext{2}{Department of Physics \& Astronomy, The University of Western Ontario, 1151 Richmond Street, London, Ontario, N6A 3K7, Canada}
\email{rahul.patel.1@stonybrook.edu}

\begin{abstract}
We present a sensitive search for \WS\ $W3$ ($12\micron$) and $W4$ ($22\micron$) excesses from warm optically thin dust around \hip\ main sequence stars within 75~pc from the Sun. We use contemporaneously measured photometry from \WS, remove sources of contamination, and derive and apply corrections to saturated fluxes to attain optimal sensitivity to $>10\micron$ excesses. We use data from the \WS\ All-Sky Survey Catalog rather than the AllWISE release, because we find that its saturated photometry is better behaved, allowing us to detect small excesses even around saturated stars in \WS.  Our new discoveries increase by 45\% the number of stars with warm dusty excesses and expand the number of known debris disks (with excess at any wavelength) within 75~pc by 29\%. We identify 220 \hip\ debris disk-host stars, 108 of which are new detections at any wavelength. We present the first measurement of a 12$\micron$ and/or 22$\micron$ excess for 10 stars with previously known cold (50--100~ K) disks. We also find five new stars with small but significant $W3$ excesses, adding to the small population of known exozodi, and we detect evidence for a $W2$ excess around HIP96562 (F2V), indicative of tenuous hot (780~K) dust. As a result of our \WS\ study, the number of debris disks with known 10--30$\micron$ excesses within 75~pc (379) has now surpassed the number of disks with known $>30\micron$ excesses (289, with 171 in common), even if the latter have been found to have a higher occurrence rate in unbiased samples.
\end{abstract}

\section{Introduction}

      Numerous surveys have been conducted to search for dusty disks around main sequence stars over the last three decades. The all-sky survey performed by the \textit{Infrared Astronomical Satellite (IRAS)} was the first to detect infrared (IR) excess emission from circumstellar dust disks at 25 and 60$\mu m$, with $\sim$~170 disks identified in all.  Subsequent pointed surveys with the \textit{Infrared Space Observatory} (\iso\ ), the \textit{Spitzer Space Telescope}, and the \textit{Herschel Space Observatory}, and the recent all-sky survey by the \akari\ satellite have greatly increased the number of disks discovered.  To date, over 350 debris disks are known around main sequence stars within 75~pc \citep[e.g.,][and references therein]{Su2006, Moor2006, Moor2009, Moor2011, Bryden2006, Rhee2007, Trilling2008, Hillenbrand2008, Carpenter2009, Mizusawa2012, Fujiwara2013, Eiroa2013, Wu2013, Cruz-SaenzdeMiera2013}, and several hundred more around more distant stars, including open cluster members, out to $\sim$1~kpc (e.g., Siegler 2007, Currie 2008a,b)
      
      Most ($\sim$85\%) of the known debris disks in the solar neighborhood are comprised of cold ($<$100~K) circumstellar dust.  These have been identified through their characteristically strong emission at wavelengths longer than 30~$\micron$, at which the disks are often orders of magnitude brighter than the stellar photosphere. This cold dust is analogous to debris produced from destructive collisions in the solar system Edgeworth-Kuiper Belt (EKB). The dust has to be continually produced in such collisions because its lifetime in the system is short: large grains spiral into the star due to Poynting-Robertson drag, and small grains are blown outward by radiation pressure. Both processes remove dust on characteristic time scale shorter than one million years \citep{Backman1993}: much less than the ages of stars in the solar neighborhood. Except in cases of stars with obvious signatures of youth, the detection of cold circumstellar dust demonstrates the presence of a belt of colliding planetesimals which, like the dust, are likely located in the cold outer reaches of the system (i.e., $>$ 10~AU from the star). 
      	
	Most known, faint warm debris disks have been discovered from pointed surveys with \spitzer\  \citep[e.g.,][]{Su2006, Trilling2008, Carpenter2009}.  Deep targeted observations with the \spitzer\  Infrared Spectrograph \citep[IRS;][]{Houck2004}, in particular, have allowed the measurement of excesses peaking in the 10--30~$\micron$ range at only 3\% of the photospheric flux at the same wavelengths \citep{Carpenter2009, Lawler2009}.  The advantage in using 5--30~$\micron$ mid-IR spectroscopy is that it allows an accurate calibration of the stellar photospheric flux---essential for detecting small excesses.  However, pointed surveys by design are limited in scope, and the data interpretation is subject to biases in the sample selection.  
  
	\WS\ offers an opportunity to search for warm debris disks over the entire sky in an unbiased fashion.  Though not as sensitive as deep, pointed Spitzer observations, \WS\ is 100--600 times more sensitive than {\it IRAS} and 10--50 times more sensitive than \akari\  in the mid-IR --- making it by far the most sensitive all-sky survey at these wavelengths. Through near-simultaneous and uniform 3--30~$\micron$  photometry, \WS\ also enables accurate calibration of the stellar photospheres, and hence good sensitivity to faint mid-IR excesses with $<10$\% of the 10--30~$\micron$ photospheric flux.

	Numerous searches of the \WS\ catalog have already been conducted to identify debris disks.  \citet{Krivov2011}, \citet{Morales2012}, \citet{Ribas2012}, \citet{Lawler2012}, and \citet{Kennedy2012} sought $W3$ and $W4$ excesses among known extrasolar planet hosts.  Approximately two dozen distinct planet-host stars with possible $W3$ or $W4$ excesses are found among these studies. \citet{Rizzuto2012}, \citet{Riaz2012},  \citet{Luhman2012}, and \citet{Dawson2013} sought \WS\ excesses in the young Scorpius-Centaurus association.  The total number of disks identified in these studies is $\approx$160, with some duplications and/or non-confirmations among the three teams (note that not all of these were debris disks).  Finally, \citet{Avenhaus2012}, \citet{Kennedy2013}, \citet{Wu2013}, \citet{Cruz-SaenzdeMiera2013}, and \citet{Vican2014} sought debris disks among solar neighborhood stars.  \citet{Avenhaus2012} find no new $W3$ or $W4$ excesses around the 100 nearest M dwarfs.  \citet{Kennedy2013} identify 15 known and 7 new $W3$ excesses around \hip\  stars within 150~pc.  An excess at such relatively short wavelengths may indicate the presence of an exozodi: a dust population at a similar temperature to the solar system's zodiacal dust.

The recent studies of \citet{Wu2013} and \citet{Cruz-SaenzdeMiera2013} are most similar to ours in design.  \citet{Wu2013} seek $W4$ excesses around \hip\ stars of all spectral types within 200 pc, while \citet{Cruz-SaenzdeMiera2013} seek $W4$ excesses around F2--K0 stars brighter than $V=15$~mag.  As we discuss in \S \ref{sec:comparison2Wu2013}  and \ref{sec:comparison2Cruz2013}, our results are mostly complementary to the results from these studies.  Importantly, through a careful calibration of \WS\ photometric systematics, we are able to detect excesses that are fainter than those reported in \citet{Wu2013} and \citet{Cruz-SaenzdeMiera2013}.  Our newly-identified disk-host stars are also often either brighter (saturated in \WS) than those considered in \citet{Wu2013} and \citet{Cruz-SaenzdeMiera2013}, or fainter (with $W4$ SNR less than 20) than those considered in \citet{Wu2013}.
	
	An accurate understanding of \WS\ photometry systematics is essential to reliable identification of dust excesses. The strongest systematic effect is the over-estimation of the $W2$ fluxes of bright ($W2<6.7$~mag) stars from profile-fit photometry \citep[see \S~VI.3.c.i.4.\ of][]{Cutri2012}, but \citet{Kennedy2013} and several additional studies also note remnant offsets in the \WS\ photometry and colors that render some previously-reported tenuous excesses uncertain.  We address this and other more subtle flux-dependent trends in the \WS\ photometry in \S~\ref{sec:wise_systematics}.  
	
	Other reasons for mis-identifications include confusion with background IR-bright sources seen in projected proximity, contamination from interstellar cirrus, and unknown amounts of interstellar extinction.  Various approaches have been adopted to mitigate these effects, including source position comparisons between the short- and long-wavelength \WS\ filters, exclusion of extended IR-bright regions in \iras, confirmation of excesses through spectral energy distribution (SED) fitting, and, importantly, visual inspection of the stellar images (e.g., Kennedy \& Wyatt 2013).  We have incorporated all of these techniques, and others, in our approach (\S\ref{sec:sample_definition}), and furthermore have only selected candidates at confidence levels greater than 99.5\% or 98\% at $W4$ or $W3$ respectively, based on the \textit{empirical} scatter in WISE photometry.  Importantly, we identify debris disk candidates using only \WS\ colors: the fact that these are homogeneous and simultaneous set of measurements reduces our vulnerability to stellar variability and other sources of error.  Our results therefore present an opportunity for an unbiased analysis of the occurrence and evolution of warm circumstellar dusty disks.  
	
	We describe the method we used to identify IR excesses in \S\ref{sec:iridentify}. We present our cross-match with the entire \hip\ catalog \citep{Perryman1997, vanLeeuwen2007} with the \WS\ All-Sky Catalog \citep[ASC;][]{Wright2010} and define our working sample of stars in \S\ref{sec:WHXD} and \S\ref{sec:sample_definition}, respectively. \S\ref{sec:atlas_vs_single} addresses a previously unknown issue that we discovered with the reliability of \WS\ ASC photometry on certain stars.  In \S\ref{sec:wise_systematics}, we outline how we precisely calibrated the \WS\ photometric systematics to produce a set of reliable debris disk detections for stars in our sample. Section \ref{sec:IRAnalysis} describes our IR excess identification procedure. Section 2.6 describes our test with identifying IR excesses in the more recent, AllWISE data release, and presents our arguments for the higher reliability of bright-star photometry in the preceding All-Sky data release. In \S\ref{sec:modeling}, we describe our procedure for quantifying basic disk characteristics.   Section \ref{sec:analysis} offers an analysis of the inferred circumstellar locations of the detected excesses: whether they belong to exozodi, asteroid belt analogs, or previously known colder EKB analogs.  Section \ref{sec:discussion} discusses our results in the context of previous surveys with {\it IRAS}, {\it Spitzer}, {\it AKARI}, and \WS.


\section{Infrared Excess Identification at $W2$, $W3$, and $W4$}
\label{sec:iridentify}

     Our goal is to determine the number of \hip\  stars with circumstellar debris disks, confined to within 75~pc of the Sun and without consideration for youth or the existence of known planets.  We search for mid-IR excesses using all \WS\ color combinations, and select stars with significant IR excesses.  Here we detail our infrared excess and debris disk candidate selection procedure.

        \subsection{\WS\  \& \hip\  Cross-match} 
        \label{sec:WHXD}

     We used the \hip\ catalog, which has photometric and parallactic measurements for 117,955 stars, as our starting sample.  We updated the stellar positions from the J1991.25 catalog epoch to J2010.54 (the mean epoch of \WS\ observations), using the \hip\  proper motions.  We positionally matched the \hip\ stars to detections from the \WS\ ASC using the NASA Infrared Science Archive cross-match service\footnote{\url{http://irsa.ipac.caltech.edu/}} and a 1$\farcs$0 matching radius.

     Following the cross-match, 159 \hip\  stars remained unmatched in \WS.  We recovered 116 of these stars in the \WS\  All-Sky Reject Table, which lists objects that were extracted from the \WS\  Atlas images, but were not included in the All-Sky Release Source Catalog because they did not meet the \WS\  Catalog source selection quality criteria \citep[see][]{Cutri2012}.  We performed this experiment only to account for unmatched \hip\  stars: we did not include objects with rejected WISE extractions in our final analysis. 
  
     The remaining 43 unmatched stars are listed in Table \ref{tab:unmatched} along with reasons for their omission. In the end, a total of 117,912 of the original 117,955 \hip\  stars were positionally matched with \WS\ sources, and no unexplained match-failures remained.

   		\subsection{Sample Definition} 
   		\label{sec:sample_definition}

	We define two samples of stars from \hip\ : a parent sample and a science sample. The parent sample consists of \hip\  stars within 120~pc of the Sun with parallaxes accurate to better than 20\%. This provides us with a large enough population of stars to determine the photospheric \WS\ color dependencies. These stars are mainly within the Local Bubble \citep{Lallement2003}, have little line-of-sight interstellar extinction ($A_V\lesssim0.5$~mag), and are suitable for correlating optical and infrared colors.  The science sample is a subset of the parent sample limited to 75~pc.  These are stars with accurate parallaxes, giving a clear volume limit to our study.  In this study we report and analyze detections of debris disks only around stars in the 75~pc science sample.
	
	For improved reliability of our debris disk-host candidate selection, we applied a number of selection criteria to the parent and science samples.  These are described in detail below, and summarized at the end of this section.
	
				\subsubsection{Parent Sample: Stars Within 120~pc}\label{sec:parent_sample}	
	
	We first eliminated stars within $5^{\circ}$ of the galactic plane. Despite angular resolution 2.5--5 times better than that of {\it IRAS} at 12 and 25$\mu m$, \WS\ images still face strong contamination from interstellar cirrus close to the plane of the Galaxy.  In addition, the local background for \WS\ photometry is estimated from a 50$\arcsec$--70$\arcsec$ annulus around each target, which can result in erroneous flux measurements when the surrounding sky brightness varies on these scales.
	
\begin{figure}
\centering
\includegraphics[scale=0.7]{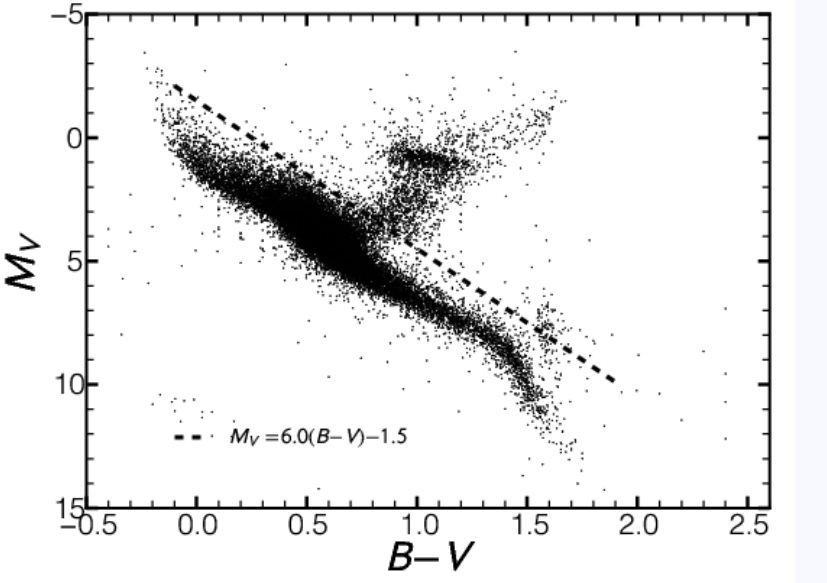}
\caption{All \hip\  stars in our $\leq$120~pc parent sample fall below the prescribed absolute magnitude cut to remove giant stars. The parent sample is restricted to stars with d~$\leq$~120~pc to reduce the effects of reddening from interstellar cirrus. Stars are also restricted to positions outside the galactic plane ($|b|\geq 5^{\circ}$) to minimize photometric contamination from confusion or interstellar cirrus.}
\label{fig:WHXD_CMD}
\end{figure}

     We further removed classes of stars in which mid-IR excesses are unlikely to be caused by circumstellar debris disks.  We followed a procedure similar to the one described in \citet{Rhee2007} to remove giant stars from our sample, by  placing an absolute magnitude restriction: we retained only stars fainter than $M_V = 6.0 (B-V) - 1.5$~mag (Fig.~\ref{fig:WHXD_CMD}).  We removed stars with SIMBAD luminosity classes of I, II or III that were missed during the color cut, and other non-main sequence stellar objects: 
post-AGB stars, 
white dwarfs, carbon stars, novae, cepheids, cataclysmic variables, high-mass x-ray binaries, planetary nebulae, and Wolf-Rayet stars. Similarly to \citet{Rhee2007}, we threw out O--B7 stars ($B_T-V_T \leq -0.17$~mag) to avoid contamination in our IR excess selection from free-free emission associated with strong stellar winds. We also removed stars redder than $B_T-V_T=1.4$~mag. These stars were removed because of the wider dispersion of photospheric \WS\ colors at late spectral types.  Some late-type (K and M) stars did possess non-photospherically blue ($B_T-V_T<1.0$~mag) colors, likely because of chromospheric activity.  A star whose $B_T-V_T$ color was $>$0.3~mag discrepant from the mean of its spectral type \citep{Pecaut2013} was assigned the mean spectral type color \citep[converted from $B-V$ using the relations in][]{Mamajek2002}.

    During the course of this study, we also discovered discrepancies in the photometry between the combined \WS\ Atlas images and the mean of the single-frame images in the $W1, W2$ and $W3$ bands. In some cases, these measurements would differ by over a magnitude. Since a definitive solution had not yet been issued by the \WS\ team at the time of this writing, we have removed from our sample stars whose ASC photometry deviates from the mean single exposure measurements by more than 2$\sigma$. Our discovery of this problem and removal of affected stars are detailed in \S~\ref{sec:atlas_vs_single}.

    We further limited our photometric candidate selection to the magnitude ranges where \WS\ photometry is reliable.  Aperture photometry is not dependable for stars brighter than $W1=8.5$~mag, $W2=6.7$~mag, $W3=3.8$~mag and $W4=-0.4$~mag.  However, \citet{Cutri2012} show that profile-fitting photometry, which relies on unsaturated pixels in the stellar halo, can consistently extract objects as bright as $W1\approx4.5$~mag and $W2\approx2.8$~mag.  We therefore apply these brighter $W1$ and $W2$ limits in our candidate selection.  In \S~\ref{sec:wise_systematics} we discuss corrections for systematics in the \WS\ photometry that are particularly pronounced for saturated point sources.   We retain the saturation levels in $W3$ and $W4$ as the brightness limits for candidate selection, since profile-fitting is not as well behaved on saturated sources in these bands.

	Finally, we applied several additional criteria that ensured good quality photometry---unconfused, uncontaminated, and with adequate SNR---including checking of the detection significance, contamination by nearby resolved companions or extended sources in \mass\ and consistent variability flagging in $W1$ and $W2$.
	
	In summary, our study samples included only stars with:

\begin{enumerate}
    \item upper limits to their \hip\  trigonometric distances that place them within 120~pc for the parent sample or within 75~pc for the science sample, and parallax accuracy better than 20\%;
    
    \item galactic latitudes $|b| > 5^{\circ}$;
    
    \item available $B_T-V_T$ colors and $\sigma_{B_T-V_T}<0.15$~mag from the \textit{Tycho-2} catalog;
    
    \item $V$-band absolute magnitudes $M_V > 6.0 (B-V) - 1.5$~mag and spectral classes excluding I, II, and III;
     
 	\item $-0.17 < B_T-V_T < 1.4$~mag and spectral type B8 or later;
    
    \item SIMBAD object descriptions excluding non-main sequence stellar objects: post-AGB stars, 
white dwarfs, carbon stars, novae, cepheids, cataclysmic variables, high-mass x-ray binaries, planetary nebulae, or Wolf-Rayet stars;
    
    \item no $\Delta K_s \leq$~5~mag projected companions within $16\arcsec$ from \mass: applied to exclude unresolved sources in \WS;
    
    \item no projected companions within $5\arcsec$ from the Visual Double Stars in \hip\ Catalog \citep{Dommanget2000}: applied to exclude unresolved sources in \WS;
    
    \item photometry that is not contaminated by known \mass\  extended sources, i.e., including only stars with \WS\ \verb|ext_flg| = 0 or 1;

	\item flux limits of $W1>4.5$~mag or $W2>2.8$~mag, corresponding to the limits of self-consistent profile-fitting photometry on saturated stars;

    \item unsaturated detections in at least one of $W3$ ($>$3.8~mag) and $W4$ ($>-0.4$~mag), with SNR~$\geq$~5; 
   
     \item \WS\ confusion flags indicative of unconfused photometry: i.e., only stars with \verb|cc_flg|[$Wi$] = 0; 
     
     \item consistent variability detections in $W1$ and $W2$, where we excluded stars whose \verb|var_flag|[$W1$] $>8$ and \verb|var_flag|[$W2$]$<5$ or \verb|var_flag|[$W1$] $<5$ and \verb|var_flag|[$W2$]$>8$. 

    \item photometry that is not severely contaminated by scattered moonlight in the $W3$ or $W4$ bands, i.e., excluding stars with \verb|moon_lev|[$Wi$]$ \geq 8$ corresponding to $>80$\% frames being contaminated by scattered moonlight in these bands;
   
   \item $W1$ or $W2$ ASC profile-fit photometry is $<2\sigma$ discrepant from the mean photometry of the All-Sky Single Exposure (L1b) Source Table. We detail this in $\S~\ref{sec:atlas_vs_single}$.

\end{enumerate}

The total number of \hip\ stars that passed criteria 1--9 was 17,499: 15\% of the full \hip\ catalog, but 63\% of all \hip\ stars within 120~pc and more than $5^{\circ}$ from the galactic plane, and 71\% of 
main-sequence stars within the $-0.17<B_T-V_T<1.4$ color range. 
Our study thus includes the majority of \hip\  main sequence stars in the solar neighborhood. 

Criteria 10--15 are band-dependent: the numbers of stars that passed all the criteria in each band with distances less than 120~pc are between 12,942 and 15,245 (Table~\ref{tab:color_analysis_summary}).  A total of 16,960 unique stars passed all our selection criteria for a sufficient subset of the \WS\ bands that we could meaningfully probe them for IR excesses at $W3$, $W4$, or (most often) both.\\

          \subsubsection{Science Sample: Stars Within 75~pc}\label{sec:science_sample}	

The science sample is further limited to stars within 75~pc, with a fractional completeness similar to that of the parent sample.  It includes 8,370 stars, constituting 67\% of \hip\ main sequence stars at $|b|>5^{\circ}$ with $-0.17<B_T-V_T<1.4$ within 75 pc.  Here also, band-dependent constraints cause the total number of stars to vary between \WS\ bands (see Table~\ref{tab:color_analysis_summary}). Since not all the stars in our science sample have valid photometry in all four \WS\ bands, we make use of all possible \WS\ color combinations to probe for excesses. Stars with debris disks reveal themselves by exhibiting anomalously red values for some subset of these colors, depending on the dust temperature --- and probing all possible colors allows us to maintain sensitivity to disks at a wide range of plausible temperatures even when one band is missing.
 
    \subsection{Discrepancy Between \WS\ Single Exposure and Atlas photometry}
    \label{sec:atlas_vs_single}
    
    Data in the ASC are created by co-adding frames from the All-Sky Single Exposure (L1b) Source Table, using the individual frame exposures acquired through each pass of the satellite in its orbit on the same part of the sky. The details of this process can be found in $\S$VI of the \WS\ All-Sky Explanatory Supplement \citep{Cutri2013a}. The mean of profile-fit photometric measurements from the Single Exposure Source Table is generally very consistent with the ASC measurements made from co-adding the same frames. 
    
    However, we have found some unexpected instances of large discrepancies between the two values, for individual objects in the $W1, W2$ and $W3$ bands. As an example, for HIP~3505, the ASC gives $W1 = 5.118\pm 0.023$~mag, but the mean magnitude measured from 13 individual exposures is $W1 = 4.3$~mag (this is after clipping any deviant individual measurements).  Similarly, $\sim$0.9~mag discrepancies exist for the $W2$ and $W3$ photometry on the same object. The \mass\ $K_s$ magnitude for this star is $4.359\pm0.016$~mag: consistent with the $W1$ mean Single Exposure measurement but not with the ASC.  We note that \citet{Mizusawa2012} did already independently conclude that the \WS\ photometry for HIP~3505 is in error.  We found similarly erroneous data for HIP 47007 and HIP 111278. All of these stars are saturated in one or more of the \WS\ bands, but the \WS\ Explanatory Supplement indicates their profile-fitting photometry should still be reliable and consistent.  The reason for these occasional discrepancies of up to $\sim$1~mag is at present unclear. For the \WS\  $W1-3$ bands, this issue affects only a tiny fraction of the photometry ($\sim$0.4\%-0.9\%); it affects $\sim$10\% of the $W4$ photometry.

	Since the goal of this study is to search for outlying photometric measurements due to debris disk emission, spurious outliers (even if rare) are a problem that must be addressed. We were faced with the choice of using mean single-exposure fluxes for our analysis, or proceeding with ASC fluxes but removing from our sample all stars with significantly discrepant ASC vs.\ mean single-exposure measurements.  We chose to retain the ASC fluxes, since in the vast majority of cases these are reliable.  However, we opted to reject from our sample all stars with $>2\sigma$ discrepancies between the two flux estimates.

 	\subsection{Correction of \WS\ Photometric Systematics on Saturated Stars} 
		\label{sec:wise_systematics}

	\WS\ photometry on faint ($11\lesssim W1\lesssim14$~mag, $9\lesssim W2\lesssim13$~mag) stars is highly consistent with \spitzer\  IRAC channels 1 and 2 photometry.  However, \citet[][\S~VI.3.c.i.4.]{Cutri2012} note that the \WS\ profile-fitting photometry on bright stars displays systematic trends when compared to the \mass\  $K_s$ magnitudes of the same stars.  The effect is strongest for saturated ($<$6.7~mag) stars in $W2$, and is present at smaller levels in $W1$.  While the photometry on saturated stars can a priori be expected to be less reliable, the \WS\ profile-fitting algorithm is designed to produce a flux estimate using the unsaturated pixels around the periphery.  Profile fitting indeed produces consistent results without increase in scatter up to 4 magnitudes beyond saturation (8.5~mag) in $W1$ (Fig.~\ref{fig:wise_flux_biases}a).  For $W2$, however, a systematic trend of flux over-estimation starts about 0.5~mag beyond saturation and continues to some of the brightest measured stars (Fig.~\ref{fig:wise_flux_biases}b).
	
	\citet{Cutri2012} illustrate the \WS\ photometric bias on bright stars using plots of the $K_s-$~\WS\ colors of $<$10~mag point sources in the \WS\ ASC. We reproduce this analysis using the B8--A9 stars in our science sample and $B-V<0.10$~mag A0 stars from the Tycho-2 Spectral Type Catalog \citep{Wright2003}. This sample of stars was chosen to reduce any shift of the $K_s-WISE$ color locus to the red. 

While most of the $K_s-Wi$ colors are close to the 0.0~mag expectation for unextincted main sequence stars of spectral type B8--A9 or earlier, we note the following effects:
 
	\begin{itemize}
	\item The $K_s-W1$ colors are systematically offset by $+0.031$~mag from zero color in unsaturated stars ($W1>8$~mag).  
	\item The $K_s-W2$ colors scatter around $-$0.004~mag for $W2>6.7$~mag; below 6.7~mag the $W2$ magnitudes are systematically over-estimated, following a well-defined trend with $W2$ magnitude up to $W2\approx2.8$~mag.
	\item In saturated stars brighter than approximately $W1=4.5$~mag or $W2=2.8$~mag the scatter in the photometry is very substantial, and there are few data points available to establish reliable trends.  We have therefore rejected from our sample all stars brighter than these limits.  
	\item There are no significant systematic trends in $K_s-W3$ or $K_s-W4$.  $W3$ photometry on saturated stars shows a large scatter, and we have excluded these altogether.  There is also an increase in scatter toward the faint end of $W4$ because the fluxes of plotted stars approach the $W4 \sim 8$~mag detection limit of \WS\ \citep{Cutri2012}.

	\end{itemize}
	
	To obtain self-consistent \WS\ colors regardless of source brightness, we correct for the biases in the $K_s-Wi$ vs.\ $Wi$ color-magnitude distributions for $W1>4.5$~mag and $W2>2.8$~mag.  We fit polynomials to the two-sigma clipped $K_s-Wi$ vs. \ $Wi$ distributions (these fits are shown in Figure~\ref{fig:wise_flux_biases}), and add the fitted values to correct the $Wi$ measurement for each star.  We subtract the respective zero-point offsets (+0.031~mag for $W1$ and $-$0.004~mag to $W2$) from the corrected saturated photometry to preserve the calibration of the \WS\ photometric system.  As an estimate of the uncertainty of the saturation corrections, we use the standard error of the residuals from the fits in 0.2~mag wide bins centered on each data point.
	
	For the remainder of the analysis, we use the corrected \WS\ $W1$ and $W2$ photometry. We do not apply corrections to the $W3$ and $W4$ photometry, which do not display systematic trends with $K_s$ magnitudes (Fig.~\ref{fig:wise_flux_biases}c,d). The $W3$ and $W4$ photometric distributions also show good agreement with \spitzer\ IRAC~8$\micron$ and MIPS~24~$\micron$ respectively for bright ($W4<9$~mag and $W3<12$~mag) point sources \citep[\S~VI.3.c.i.\ of][]{Cutri2012}.

     \subsection{Debris Disk Candidate Selection}
     \label{sec:IRAnalysis}

	We identified debris disk-host candidates by selecting stars with the reddest infrared colors in color-color diagrams.  Excesses were sought in the $W2$, $W3$, and $W4$ passbands, so our analysis is sensitive to stars with excesses between 4--28~$\micron$.  The excesses were identified based purely on the \WS\ colors, without relying on photospheric fits to the spectral energy distributions.  If a star displayed a significant excess in any of the six \WS\ color combinations, it was considered a debris disk candidate.  SED fits were used at a later stage to confirm the validity of debris disk candidate identifications, and to determine the dust temperatures of high-probability debris disks.
	
	The photospheric colors of main sequence stars vary over the \WS\ bands, mostly as a function of stellar effective temperature.  We calibrated this dependence to avoid mistaking stars with intrinsically red \WS\ colors for debris disks (Fig.~\ref{fig:Res}).  $B_T-V_T$ color measurements exist for all our sample stars by design, and  are not biased by the presence of debris disks.   We used a trimmed mean to determine the mean locus of the $Wi-Wj$ vs.\ $B_T-V_T$ relations from the parent sample.  We iteratively removed the largest $Wi-Wj$ color outlier in 0.1~mag wide $B_T-V_T$ color bins until half of the data points in the bin were rejected, leaving only the data clustered near the mode of the bin.  This removed the dependence of the relation on outliers, most notably mid-IR-excess debris disk hosts.  We traced the $Wi-Wj$ vs.\ $B_T-V_T$ relations in step sizes of 0.02~mag in $B_T-V_T$.  We refer to the mean $Wi-Wj$ corresponding to a given $B_T-V_T$ color as $W_{ij}(B_T-V_T)$.  Table~\ref{tab:wise_bv_trends} lists the $W_{ij}(B_T-V_T)$ trimmed mean and its standard error (based on the surviving 50\% of data points) for all \WS\ color combinations.

	We are now in position to determine whether the \WS\ colors of any particular star reveal a significant excess.  We define the excess $E[Wi-Wj]$ in the $Wi-Wj$ color of a star with a given value of $B_T-V_T$ as:
	\begin{equation}
        \label{eq:colorexcess}
	E[Wi-Wj]=Wi-Wj - W_{ij}(B_T-V_T)
	\end{equation}
\noindent	
We then define the SNR of the excess as the ratio of $E[Wi-Wj]$ to the uncertainty $\sigma_{ij}$, 
    \begin{equation}
    \label{eq:colorsnr}
    \Sigma_{E[Wi-Wj]} = \frac{E[Wi-Wj]}{\sigma_{ij}} = \frac{Wi-Wj - W_{ij}(B_T-V_T)}{\sigma_{ij}},
    \end{equation}
\noindent where $\sigma_{ij}$ combines the $Wi$ and $Wj$ photometric uncertainties, and the standard error on $W_{ij}(B_T-V_T)$:
       \begin{equation}
       \label{eq:error}
       \sigma_{ij}  = \sqrt{\sigma_{Wi}^2 + \sigma_{Wj}^2 + \sigma_{W_{ij}}^2}
       \end{equation} 

   \noindent For shorthand, we use \ES\ throughout the rest of the paper when the discussion does not refer to any specific color.  \ES\ is plotted against $B_T-V_T$ for each color in the bottom halves of the panels in Figure~\ref{fig:Res}. 
   
	Figure~\ref{fig:colordist} shows the \ES\ distributions for each set of \WS\ colors with solid histograms.  The distributions are characterized by sharp cores and long tails to higher SNRs.  The cores of the histograms represent the random scatter around zero excess (black data points in the lower halves of the panels of Fig.~\ref{fig:Res}), corresponding to measurement and calibration uncertainties.  We estimate the rate of low-SNR false-positive excesses by mirroring (dashed histograms) the distribution of negative excesses into the positive wing.  We thus empirically construct a distribution that represents the measurement uncertainties, both random and systematic.
	
	Using the empirically determined uncertainty distribution, we can calculate the false-positive rate (FPR) for detecting excesses as a function of the threshold beyond which red outliers are designated as bona fide excesses.  The FPR is simply the number of outliers beyond the threshold in the uncertainty distribution divided by the number of red excesses beyond the threshold. For example, based on the histogram of our $W1-W4$ uncertainty distribution (see top left panel of Fig.~\ref{fig:Res}), we expect only 2 false positives beyond our chosen threshold of $\Sigma_{E[W1-W4]} = 3.19$ (vertical dashed line in the figure).  As there are 429 excesses in the actual $W1-W4$ color distribution redwards of the same limit, the empirical FPR is $2/429=0.47\%$.  Choosing a lower threshold for excess identification would produce more excesses but would increase the FPR, while choosing a higher threshold would reduce the FPR further.  Our objective in general is to obtain FPR $<$ 0.5\%.  
	
	Empirically, however, we can not determine the FPR beyond the threshold value at which the number of false positives drops to zero.  This sets an upper bound to our ability to empirically set the confidence level for excess identification.  For color distributions involving $W4$ this upper bound is between 99.8\%--99.9\%.  However, the $W1-W2$, $W1-W3$, and $W2-W3$ distributions do not possess $>$200 excesses with even a single false positive (such that FPR $\leq0.5\%$) at any value for \ES.  Our empirical confidence level for the $W1-W3$ and $W2-W3$ excess selection in $\gtrsim$98\%, and for $W1-W2$ it is $\gtrsim$95\%.
	
	The 99.5\% threshold that we employ for $W4$ excess selection is similar to a Gaussian 3$\sigma$ (99.8\%; one-tailed) threshold.  Importantly, however, our 99.5\% confidence threshold does not assume Gaussian error statistics: only that the distribution of uncertainties is symmetric around zero.  In addition, it includes both random and systematic errors.

	 We denote the minimum excess SNR \ES\ at the 99.5\% confidence level as $\Sigma_{E_{\rm CL}}=\Sigma_{E_{99.5}}$.  The $\Sigma_{E_{99.5}}$ threshold is between 3.16--3.26 for the three \WS\ color distributions that use $W4$ (Table~\ref{tab:color_analysis_summary}).  This compares to 2.58 for a Gaussian distribution at the 99.5\% (one-sided) confidence level.  The discrepancy is relatively small, and indicates that our corrections to the $W1$ and $W2$ saturation systematics, and to the $Wi-W4$ dependencies on $B_T-V_T$, have left us with well-behaved uncertainty distributions for the $Wi-W4$ colors.  The $\Sigma_{E_{\rm CL}}=\Sigma_{E_{98}}$ thresholds for $W1-W3$ and $W2-W3$, and the $\Sigma_{E_{\rm CL}}=\Sigma_{E_{95}}$ threshold for $W1-W2$ are also listed in Table~\ref{tab:color_analysis_summary}.  All $\Sigma_{E_{\rm CL}}$ thresholds are marked with vertical dotted lines in Figure~\ref{fig:colordist}.  With the exception of the $W2-W3$ $\Sigma_{E_{98}}$ threshold, the close correspondence between the empirical confidence levels at the various thresholds and the expectations from a Gaussian distribution, suggest that we have calibrated away systematics to the point where the uncertainty distibutions can be explained almost entirely by the random photometric errors.
	 	
	 We identified 243 stars with significant excesses within 75~pc of the Sun, the vast majority (231) of which are in $W4$.  Among which we expect only $0.5\% \times 231=1.2$ false excesses.  However, IR excesses can in principle be caused by contamination from other IR sources in the \WS\ beam (mainly IR cirrus and unresolved late-type binary companions) rather than circumstellar dust. We screen our excesses for these types of contamination, and eliminate 23 of them (mostly due to line-of-sight IR cirrus visible in the \WS\ images), leaving 220 candidate debris disks with excesses at $W2$, $W3$, or $W4$ within 75~pc of the Sun.

	 	 A summary of the number of identified mid-IR excesses, contaminated sources, and candidate debris disks for each color selection criterion is given in Table \ref{tab:color_analysis_summary}. Stars that were rejected after being identified as candidate debris disk hosts are listed in Table~\ref{tab:rejects}. The host star properties of all our identified debris disk systems are shown in Table~\ref{tab:stellarparameters}.  Table~\ref{tab:excessstats} lists the information on the significance of the excess \ES\ for each color. Since debris disk-bearing stars often have an excess in multiple \WS\ color combinations, a six character flag indicating the color excess each star has also been provided. The dust properties determined from SED fitting (\S~\ref{sec:modeling}) are given in Table~\ref{tab:diskcandidates}.

          \subsection{All-Sky vs. AllWISE Data Release}
          \label{sec:asc_v_awr}
Since the inception of this study, \WS\ has released an updated version of the all-sky survey, called the AllWISE Data Release\footnote{\url{http://wise2.ipac.caltech.edu/docs/release/allwise/expsup/index.html}} (AWR). The AWR incorporates data products taken during the NEOWISE Post-Cryo phase of the mission, and is a significant improvement over the \WS\ ASC. We incorporated the \WS\ AWR into our IR excess search in an attempt at more reliable debris disk identification.

However, we identified two issues that make the AWR less suitable than the ASC for precise identification of IR excesses. First, the $W1$ and $W2$ AWR photometry behaves less well in the saturated regimes of these bands. In particular, we find that the behavior of the $K_s-WISE$ vs. \WS\ relations for saturated $W1$ and $W2$ AWR photometry is not monotonic, unlike in the ASC. This is indeed seen in Figures 10a-b in \S~II.1.d.i of the AWR explanatory supplement, which compares the ASC data to the AWR for $W1$ and $W2$. Consistent with these observations, the AWR explanatory supplement states that ``The \WS\ ASC may provide better photometry than in the AWR for objects brighter than [$W1<8$~mag and $W2<7$~mag].'' Therefore, we abandon using the AWR $W1$ and $W2$ photometry for our analysis.

We noticed a similar issue when we attempted to identify excesses using only $W3-W4$ colors constructed from the AWR data products. Here, we found more stars with negative $\Sigma_{E[W3-W4]}$ values, that widened the \ES distribution and pushed the 99.5\% confidence threshold for $W4$ excesses to $\Sigma_{E[W3-W4]_{CL}}=9.4$. This is in stark contrast with the much tighter distribution we found using the ASC data ($\Sigma_{E[W3-W4]_{CL}}=3.2$). After closer inspection of the negative \ES valued stars, we found that the AWR $W3$ photometry was intrinsically brighter than the same ASC photometric measurement for the same star. HIP~51933 is one such example, where its AWR $W3$ profile fit photometric measurement is 0.25~mag brighter than the corresponding ASC photometry. This intrinsic brightening is seen in the majority of our negative \ES stars. We can see similar brightening of the AWR $W3$ photometry relative to the ASC in Figure 10c in \S~II.1.d.i of the AWR explanatory supplement between AllWISE $W3$ magnitude at $7.5<W3<9$~mag. The surplus of stars with negative \ES incurs a non-Gaussian component to the $\Sigma_{E[W3-W4]}$ distribution, and makes the AWR $W3$ photometry less reliable in searching for IR excesses.
 
Hence, after performing the same set of procedures outlined in \S\S\ \ref{sec:iridentify} -- \ref{sec:IRAnalysis} on data from the AWR, we determined that the ASC data are better suited for identifying IR excesses through the method outlined in the preceeding sections.

\section{Debris Disk Brightness and Temperature Determination}
\label{sec:modeling}

	We fit the photometry of our debris disk candidates using model photospheres for the stellar contribution and single-temperature blackbodies for the dust. To constrain the photospheric fits, we use optical $B \mbox{ \& } V$ Johnson photometry taken from the \hip\ catalog, $JHK_s$ photometry from \mass, $W1$, and in the lack of significant excesses ($\Sigma_E<\Sigma_{E_{\rm CL}}$), also $W2$ and $W3$ photometry from \WS.  The photometry was converted from magnitudes to $\mbox{erg s}^{-1} \mbox{cm}^{-2} \text{\AA}^{-1}$ using the Johnson, \mass\ and \WS\ zero-point fluxes \citep{Johnson1953, Cohen2003, Wright2010}. The isophotal wavelength was adopted as the central wavelength for each bandpass.
   
     We used NextGen \citep{Hauschildt1999} photospheric models for stars of A--K spectral types, and \citet{Kurucz1993} models for the few late-B stars in our candidate list. The models were fit to the calculated integrated fluxes over the bandpasses using $\chi^2$ minimization with {\sc mpfit} \citep{Markwardt2009}. The photospheric temperature ($T_\ast$), and flux scaling (i.e., stellar radius) were kept as free parameters. The surface gravity was kept constant at empirically determined values for main sequence stars from \citet{Schmidt-Kaler1982}.\footnote{Available on-line at the STScI Calibration Database System, http://www.stsci.edu/hst/observatory/cdbs/castelli\_kurucz\_atlas.html.}  
     
	In some cases our fits produced poor matches to the stellar photosphere ($\chi^2>4$). In each of these cases, the \mass\ measurements were systematically offset compared to \WS\ $W1$ and $W2$.  In such situations we used only $W1$ and $W2$ to fit the Raleigh-Jeans tail of the stellar photosphere; the stellar temperature was estimated from the SIMBAD spectral type listing and comparing it to table 5 from \citet{Pecaut2013}.

We calculate the dust excess fluxes in each \WS\ band by subtracting the photospheric flux integrated over that band ($F_\ast(\lambda_{iso})$) from the measured values ($F_{obs}(\lambda_{iso})$), thereby obtaining a value for the dust flux at $\lambda_{iso}$, the isophotal wavelenght of the band in question:

        \begin{equation}
        \label{eq:dustflux_uncor}
            F^o_d(\lambda_{iso}) = F_{obs}(\lambda_{iso}) - F_\ast(\lambda_{iso}).
        \end{equation}

	Where a significant excess is detected in both $W3$ and $W4$, we fit the measured flux excesses using a single-temperature ($T_{BB}$) blackbody model of the dust.  While the dust is not expected to be actually concentrated in a thin ring at uniform temperature and radius from the star, the calculated temperature
and circumstellar radius constitute useful estimates of the debris disk’s average properties.

Most of our excess detections are at $W4$ only. In these cases, we use the upper limit on the $W3$ excess flux to set a 3$\sigma$ upper limit on the  dust temperature. In many of these cases, the $W3$ excess, though formally insignificant, is positive.  We use these marginal $W3$ excesses to calculate a unique temperature for the dust, in addition to the upper limit already mentioned.  The data in these cases are formally consistent with arbitrarily low temperatures, but nevertheless the calculated temperature is of some value, especially when the $W3$ excess has a significance more than 2$\sigma$ and is only just below our threshold.  Both the calculated and upper-limit temperatures are given in Table~\ref{tab:diskcandidates}, and the reader should bear in mind that only the latter are guaranteed to be physically meaningful.

        We proceed in an exactly analogous way for the few disks where we have significant detections only in $W3$.  Here, we use upper limits on the $W4$ flux to set 3$\sigma$ \textit{lower} limits on the temperatures.  In every case, the nominal $W4$ excess is positive though not significant.  Thus, just as for the $W4$-only excesses with positive non-signficant $W3$ excesses, we calculate unique temperatures in addition to the limits.  These values and the limits are given in Table~\ref{tab:diskcandidates}.

In addition to dust temperatures, we derive and tabulate the values of $f_d$, the ratio of the bolometric luminosity of the dust to that of the star --- and also the circumstellar radii corresponding to dust temperatures.  We will now describe how we use measured flux excesses (or limits) in $W4$ and $W3$, obtained using Equation \ref{eq:dustflux_uncor}, to calculate the dust temperature (or limit), the value of $f_d$, and the circumstellar radius of the dust (or limit thereon).

The \WS\ magnitude-to-flux conversion assumes that the spectral slope of the excess is akin to a Vega-like spectrum (i.e., a Rayleigh-Jeans slope) at the \WS\ wavelengths.  The excess monochromatic flux from Equation \ref{eq:dustflux_uncor} therefore needs to be  color-corrected for the response of \WS\ to an emission from a cool blackbody source:

		\begin{equation}
		\label{eq:kcor}
		F_d(\lambda_{iso}) = \frac{F^o_d(\lambda_{iso})}{f_c(Wi;T_{BB})},
		\end{equation}
		
	\noindent where $f_c(Wi;T_{BB})$ are the flux correction factors like those found in Table~6 in \S IV.3.g.vi of the \WS\ Explanatory Supplement.  We have duplicated the calculations that produced these and created a lookup table of $f_c(Wi;T_{BB})$ that spans a wider and much more finely-sampled range of temperatures than that in the Explanatory Supplement. 

Since we do not know a priori the temperature of the dust, we use this lookup table to perform a grid search to find the blackbody temperature that matches our observed fluxes.  This gives us the spectrum of the dust.  As we already have the photospheric model of the star, the bolometric luminosity ratio $f_d$ may easily be found:
   
    \begin{equation}
    \label{fraclum}
    f_d = \frac{\int \! F_{\lambda,d} \, \mathrm{d}\lambda}{\int \! F_{\lambda,\ast} \, \mathrm{d}\lambda}.
    \end{equation}
    
    \noindent The disk radius is then calculated assuming that the dust ring is in thermal equilibrium with the stellar radiation:
        
        \begin{equation}\label{eq:diskradius}
            R_{BB} = (278.3/T_{BB})^2 \sqrt{L_\ast} \mbox{ (AU) }. 
        \end{equation}

Where one of the fluxes is an upper limit, the temperature will also be a limit (upper limit for a $W4$-only excess; lower limit for a $W3$-only excess).  A temperature limit converts easily into a limit on $R_{BB}$, but not into a limit on $f_d$: in general, the value of $f_d$ obtained using the equations above in the case where one of the fluxes is an upper limit will be neither the lowest nor the highest value of $f_d$ permitted by the data.  

However, we can set a meaningful lower limit on $f_d$ in every case of single-band excess. This is because the lowest value of $f_d$ consistent with the data corresponds to the case where the largest possible fraction of the disk luminosity comes out in the one band we have measured --- in other words, where the blackbody emission peaks at the band's isophotal wavelength.  This corresponds to a temperature of 131~K in the case of $W4$-only excesses or 272~K for $W3$-only excesses.  We can therefore adopt as our dust model a blackbody having whichever of these temperatures is appropriate, normalized to match the measured excess in the relevant band. Equation \ref{fraclum} then gives the \textit{minimum} $f_d$ that is consistent with the data.  This limit is given in Table~\ref{tab:diskcandidates} for all of our single-band excesses.

For some $W4$-only excesses, the $W3$ flux measurement fails to pass our selection criteria.  For these, we cannot place any constraints on the dust temperature, but we can still place a lower limit on $f_d$ as described in the preceding paragraph.  For these cases, the temperature given in Table~\ref{tab:diskcandidates} is the one corresponding to the lower-limit $f_d$ (131~K) and has no independent physical meaning.

For disks with excesses at both $W3$ and $W4$, Table~\ref{tab:diskcandidates} gives values for the dust temperature, its circumstellar radius, and its bolometric flux fraction $f_d$.  For single-band disks, the table gives limiting values for all these quantities, as well as tentative calculated values in cases where the formally non-detected band showed a positive though non-significant excess.  The SEDs of all stars with \WS\ $W3$ or $W4$ excesses, including our blackbody fits to the dust emission, are plotted in Figure \ref{fig:SED}.


\section{Analysis of Excesses and Location of the Dust}
\label{sec:analysis}

    We divide the analysis of our candidate debris disks according to the wavelengths at which they were detected. We first discuss our $W4$-only detections, which in most cases represent the short-wavelength tail of blackbody emission from cold dust peaking at longer wavelengths, although in a few cases we find evidence of multi-temperature dust. We then discuss detections of excesses at both $W3$ and $W4$ bands that may be explained by warm dust alone. Finally, we discuss the likelihood of hot dust orbiting a few stars that show significant excesses at $W2$. 
  
\subsection{$W4$-Only Excesses: Kuiper Belt Analogs and Multi-Temperature Dust Disks}
     
 	Stars with dust emission detected at $W4$, but not in any of the three colors that do not include $W4$, make up 96\% of our total detections, or 211 of 220.  Of these 211 stars, just over 50\% have been previously published as excess detections, and 36\% have published dust temperatures, mostly based on IR excess measurements at multiple wavelenths including $\lambda \gtrsim 60$ \micron.  None exhibits an excess detected at shorter wavelengths comparable to the $W3$ band (12 \micron).

        However, the dust in these systems must necessarily emit some flux at shorter wavelengths, even though it is not above our $W3$ detection threshold.  The existence of such flux, undetectable from any individual star, can nonetheless be divined from the distributions of $E[W1-W3]$ and  $E[W2-W3]$ (defined in Equation \ref{eq:colorexcess}).  If there were no $W3$ flux from the dust, these distributions would be symmetric around zero, with the numbers of positive and negative values equal to within statistical uncertainties.  Instead, we find that they are strongly skewed toward positive values. This observation suggests that we can measure the $W3$ excess flux, in aggregate, for these nominally $W4$-only systems.  Such measurements allow us to determine the averaged dust temperature of various subsets of the $W4$-only systems, even though only an upper limit can be placed on the temperature of each dust-disk individually.

        Because the distances and dust-luminosities of stars in our sample vary widely, we perform such analyses by calculating the $W3/W4$ excess flux ratios, rather than simply the $W3$ excess flux. We have a $W3$ measurement that meets the selection criteria given in \S \ref{sec:sample_definition} for 183 of our 211 $W4$-only detections.  
The weighted mean of the uncorrected $W3/W4$ flux ratio for all 183 stars is $0.174 \pm 0.026$.  Thus we have a highly significant detection of the aggregate $W3$ excess, even though none of these stars had individual $W3$ excesses above our detection threshold.  This calculation can be repeated for specific subsets of these 183 stars, with interesting implications for the characteristic dust temperatures.  We perform these calculations below in \S \ref{sec:W4prev} and \S \ref{sec:W4new}.  

\subsubsection{$W4$-Only Excesses with Prior Longer Wavelength Detections} \label{sec:W4prev}

        Of our 183 stars with $W4$-only excesses and $W3$ fluxes passing our selection criteria, 95 were previously known to exhibit IR flux excess, in many cases due to measurements at wavelengths longer than 30 \micron.  Of these 95 stars, 46 have published dust temperatures below 130~K, 20 have published dust temperatures of 130~K or higher, and 29 have no previously published dust temperatures.  For convenience, in this section we will refer to these three samples of stars as the `known cold disks', the `known warm disks', and the `published disks of unknown temperature'.  

        The published dust temperatures of the 46 known cold disks, by construction, all correspond to dust colder than the asteroid belt in our own Solar System.  They range down to 50~K, just slightly warmer than the Solar System's EKB.  For these 46 stars, we find an aggregate $W3/W4$ excess flux ratio of $0.122 \pm 0.028$.  The fact that this ratio is not statistically consistent with zero means that we have detected a statistically significant $W3$ excess in the aggregate of these systems, though not in any one individually.  This is the first indication of excess flux at wavelengths shorter than 18 \micron\ for any of these systems.

        We convert this aggregate $W3/W4$ excess flux ratio to a blackbody temperature, which will approximate the flux-weighted mean temperature of dust in the known cold disks.  The correction factors $f_c(Wi;T_{BB})$ must be taken into account in this conversion, and we do not know their values a priori since they depend on the temperature we seek to determine. Since it is easy to solve the inverse problem of predicting the uncorrected $W3/W4$ excess flux ratio for dust at a given blackbody temperature, we perform the conversion by a simple grid search in temperature space, finding that the uncorrected excess flux ratio $W3/W4 = 0.122 \pm 0.028$ corresponds to a blackbody temperature of $90 \pm 6$~K.  For comparison, the median published dust temperature for these disks is 85~K (see \S~\ref{sec:discussion} and Figure~\ref{fig:Venn} references).  Our $90 \pm 6$~K aggregate temperature, which was measured using shorter wavelengths than any of the published temperatures, is consistent with this result: it appears that at $W4$ and $W3$, we are measuring the Wien tail of blackbody emission from the same cold dust seen at longer wavelengths.

        The known warm disks have published temperatures ranging from 130~K to 276~K \citep[with one outlier at 1700K;][]{Matranga2010}.  This dust could be analogous to the asteroid belt and even the zodiacal dust in our Solar System.  Our aggregate $W3/W4$ excess flux ratio from these 20 stars is $0.68 \pm 0.21$.  This much higher result relative to the known cold disks is expected given that the warm dust will emit more at shorter wavelengths.  Our $W3/W4$ excess flux ratio corresponds to an aggregate dust temperature of $154 \pm 19$~K.  This is consistent with the median published dust temperature of 178~K for these disks,corresponding to a disk brightness of $f_d = 3.93\times10^{-5}$. This aggregate temperature also indicates a weak contribution from any exo-zodi (300~K) dust emission in these systems. We calculate the contribution of any such exo-zodiacal dust in the aggregate by assuming the $W3$ excess aggregate flux is arises from 300~K dust. Using the $2\sigma$ upper limit on the $W3$ excess aggregate flux, we calculate an upper limit dust brightness $f_d = 2.48\times 10^{-5}$. This is 37\% smaller than the actual disk brightness for the aggregate. Consequently, the $W4$ excess produced from this dust emission is 80\% fainter than that of the derived aggregate, evidence of non exozodii dust emission in the aggregate.
        
        For the 29 previously published disks of unknown temperature, we find an aggregate $W3/W4$ excess flux ratio of $0.30 \pm 0.14$.  As this value is too uncertain to be useful, we combine the published disks of unknown temperature with our own newly discovered disks in \S \ref{sec:W4new} below.

\subsubsection{New $W4$-Only Excesses} \label{sec:W4new}

        Of our 183 stars with $W4$-only excesses and $W3$ fluxes passing our selection criteria, 88 have not been previously published as IR excesses at any wavelength.  These excesses are too tenuous ($<$10\%) to have been accurately measured with \iras\ or \akari, and the stars have not been targeted with \spitzer\ or {\it Herschel}.  They have not been identified as excesses in previous analyses of the \WS\ data.

        Calculating the aggregate $W3/W4$ excess flux ratio is of particular importance for these systems, because if the systems correspond to real dust disks at physically plausible temperatures, a detectable aggregate $W3$ excess must be present. Lack of such a detection would falsify the $W4$ excesses, suggesting that they were due to imperfectly understood systematics in $W4$ rather than to genuine dusty disks.

        The aggregate $W3/W4$ excess flux ratio for these is $0.508 \pm 0.082$, corresponding to a highly significant detection of the aggregate $W3$ excess flux.  This ratio maps to an aggregate temperature of $139 \pm 8$~K.  These significant, consistent, and physically reasonable results constitute a useful check, and confirm that our new $W4$-only excesses are real dust disks not identified by previous studies.

        We can also add the sample of previously published disks of unknown temperature, mentioned in \S \ref{sec:W4prev} above, to the sample of 88 new disks, and calculate the aggregate ratio of the combined samples.  This is interesting because most of the 29 previously published disks of unknown temperature were also identified using \WS\, and thus the result will yield an estimate of the characteristic dust temperature of disks that were not detected in previous surveys (ISO, IRAS, AKARI), but have recently been identified using \WS.  The aggregate $W3/W4$ excess flux ratio for this combined sample of 117 disks is $0.458 \pm 0.071$, which corresponds to a temperature of $134 \pm 8$~K.  This temperature is comparable to the outer edge of our own asteroid belt.

\subsubsection{Summary}
       We have found conclusive evidence for an aggregate $W3$ excess from stars that individually have significant excesses only at $W4$.  Known cold disks have aggregate $W3/W4$ excess flux ratios implying cold dust and known warm disks have aggregate excess flux ratios consistent with warm dust.  Disks recently discovered in this work and other studies using \WS\ $W4$ photometry show intermediate flux ratios that correspond, interestingly, to the temperature of dust located near the frost line and emitting its peak blackbody flux in the $W4$ bandpass.  This aggregate temperature is only the mean of a potentially very wide distribution, but it is nonetheless possible that most of the newly discovered disks are warm (i.e. $>100$ K): if the $W4$ excesses measured for these systems were all merely the Wien tails of cold-dust emission, the cold dust in at least some cases would likely have already have been detected at 60 \micron\ by {\it IRAS}.


	\subsection{$W3$ and $W4$ Excesses: Asteroid Belts and Exozodi}
	\label{sec:W3andW4excesses}
	
	We find four stars with significant excesses in both $W3$ and $W4$ but not in $W2$: HIP~7345 (49~Cet), HIP~24528 (HD~34324), HIP~41081 (HD~71043) and HIP~95261 ($\eta$~Tel).
	Their blackbody dust temperatures can be determined exactly and reliably, and are given in bold in Table~\ref{tab:diskcandidates}.  All of these are known debris disk-host stars with 24$\micron$ excesses from \spitzer, 25$\micron$ excesses from \iras, or 22$\micron$ excesses from the recent \WS\ study by \citet{Wu2013}, and with longer-wavelength detections at either 60$\micron$ (\iras), or 70$\micron$ (\spitzer).  Their published dust temperatures based on the longer-wavelength results range 80~K to 150~K. Our measured dust temperatures are higher in every case, ranging from 133~K to 199~K.  These temperatures are well-matched to the 130--190~K temperature range corresponding to the asteroid belt in our own Solar System; by contrast, the published temperatures mostly correspond to dust much colder than our asteroid belt, though not at the 30--55~K temperatures characteristic of Solar System Kuiper Belt objects.

        The discrepancies between our dust temperatures for these objects and the published ones based on longer-wavelength excesses demonstrates the existance of dust at multiple temperatures.  HIP 95261 has the lowest discrepancy (177~K vs. 150~K) and HIP 41081 the greatest (199~K vs. 91~K). Even for HIP 95261, the discrepancy is likely real and points to a dust distribution spaning a wide range in circumstellar radius. The much larger discrepancy seen for HIP 41081 could even indicate two distinct dust populations at different radii and temperatures, separated by a gap --- however, detailed modeling to distinguish this possibility from a single dust distribution spanning a wide range in circumstellar radius and temperature is beyond the scope of this work.  In any case, all of these objects are extremely interesting as targets for further study and observations, both to map the dust in more detail and to search for possible associated planets.
 
     We also find five stars with excesses that are significant only at $W3$: HIP~19610, HIP~51793, HIP~80781, HIP~102238 and HIP~109656.  All are new discoveries of our survey, with no previously published IR excess detection at any wavelength.  All five have positive though formally non-significant $W4$ excesses, a statistical result which strongly suggests that the dust is emitting flux at $W4$, even though it is below our detection threshold.  

We use upper limits on the $W4$ excess in these systems to calculate $3\sigma$ lower limits on the temperatures.  These range from 174~K (HIP~80781) to 274~K (HIP~19610), although we caution that for HIP~80781 and HIP~109656 the $W4$ fluxes are suspect due to the discrepancy between the ASC and single-exposure photometry discussed in \S \ref{sec:atlas_vs_single}, and were therefore not used in our search for excesses within the science sample.  Nevertheless, the fluxes may be accurate for these objects, and certainly are for the other three stars.  Thus our 3$\sigma$ lower limits on the dust temperatures conclusively demonstrate (at least for the three stars with good $W4$ photometry) that we are not merely measuring the Wien tail of blackbody emission from cold dust. Rather, dust exists at asteroidal (130--190~K) or, more likely, even warmer temperatures in these systems.  

It is highly likely that the dust in these systems overlaps the habitable zone, which corresponds to temperatures of 230--330~K.  This dust is likely produced by mutual collisions between asteroidal objects warmer and far more abundant than those in our Solar System --- objects that could be leftovers from the formation of one or more potentially habitable planets.  Interestingly, however, the lack of significant excess detections at wavelengths greater than 12 \micron\ suggests there is no Kuiper Belt analog in these systems, and therefore the overall system architecture may be very different from that of our own Solar System.  Such systems could serve as a probe of the diverse evolutionary pathways the process of planet formation can follow.


	\subsection{$W2$ Excesses: Hot Dust or Signs of Chromospheric Activity}\label{sec:W2Excess}

	Our $W3$ and $W4$ analyses are naturally extendable to $W2$, and we sought hot-dust excesses from the $W1-W2$ color distribution. We found eight stars within 75~pc with significant $W2$ excesses.  As discussed in \S\ref{sec:IRAnalysis}, our empirical calibration of false positives does not allow us to push our confidence threshold beyond 95\% for the $W1-W2$ excesses.  Nonetheless, this still implies that among the eight $W2$ excesses we expect less than one to be caused by random error. 
	
	We exclude two of the excesses from further consideration, as they are associated with unresolved binary stars with disparate spectral types: HIP~999 (G8V+K5; composite spectral type of K0 in \hip) and HIP~3121 (K5V+M3V).  That is, in these two cases an inaccurate estimate of the joint photospheric $W1-W2$ color of the binaries is indeed the likely cause for the small $W2$ excesses.  This conclusion is supported by the fact that these stars also possess small, sometimes significant, $W1-W3$ and $W1-W4$ excesses: that is, a blackbody slightly cooler than the $BVJHK_sW1$ photospheric fit---the secondary component---is needed to explain the \WS\ SED.  A third $W2$ excess star, HIP~3729 (K2Ve), is a suspected double-lined spectroscopic binary, although according to \citet{Torres2006} that classification is uncertain because of the star's large $v\sin i$ (75~km~s$^{-1}$).  We observe that this star shows marginal excesses at all \WS\ wavelengths, including $W1$: a signature of variability between the \mass\ and \WS\ epochs, rather than a bona-fide excess.  It is possible that the \WS\ excesses are caused by geometric factors affecting the combined flux from an unresolved close binary: e.g., grazing eclipses or ellipsoidal variations.  Therefore, we also exclude HIP~3729.
	
	The remaining five stars are not known to be in binary systems: HIP~30893 (K2V), HIP~74235 (K2V), HIP~74926 (K5Vp), HIP~96562 (F2V), and HIP~109941 (K5V).  Their SEDs stars are shown in Figure~\ref{fig:SEDw2}.  
	Four of the five stars show small, sometimes significant $W1-W3$ and $W1-W4$ excesses (Table~\ref{tab:excessstats}), and for three of them the $W1$ data point is also marginally above the fitted photosphere.  Previously unknown close companions could account for these, in much the same way as for HIP~999, HIP~3121, and HIP~3729.  However, being within 75~pc and relatively cool, these stars have been prime targets for radial velocity monitoring and planet searches.  Therefore, we assume that the excesses from these four stars are not caused by unknown stellar companions.  The remaining $W2$ excess star, HIP~74235 (K2V), exhibits no excess at any other wavelength.  All of its non-$W1-W2$ excesses are negative---most marginally, except for $W2-W3$---indicating that the apparent excess is localized to the $W2$ band.

	A potential clue to the nature of the detected $W2$ excesses is the fact that four of the five stars have K spectral types, and only one is hotter (F type).  This may suggest that an inaccurate photospheric correction of the $W1-W2$ color may be to blame for the large fraction of K-star $W2$ excesses in our science (75~pc) sample.  However, the larger parent (120~pc) sample selection also contains A through G-type $W2$-excess stars, with no additional $W2$ excesses from K stars.  This is evident from the distribution of $W1-W2$ excesses as a function of $B_T-V_T$ in the bottom right panel of Figure~\ref{fig:Res}: the $W1-W2$ excesses do not cluster at red $B_T-V_T$ colors.  The dominance of K star excesses in the 75~pc sample may therefore be attributable to the higher photometric precision that can be attained on faint K dwarfs near the Sun.  We conclude that these excesses are real.
	
		All five of the detected $W2$-excess systems may possess small amounts of hot dust, between $\sim$400~K--900~K.  Such dust would be in close proximity to the star, and would be expected to be very short-lived: potentially indicative of the recent planetesimal activity in the innermost reaches of these systems.  The excess from the one F star (HIP~96562) is fully consistent with a $T_{\rm eff}=780$~K black body.  The remainder of the excesses, around the four K stars, require steeper than Raleigh-Jeans SEDs to fit the lower $W3$ and $W4$ excesses.  Such SEDs would be representative of sub-micron dust grains with low emissivity at $>$5\um\ wavelengths.  We use modified blackbodies to model these:
    \begin{equation}\label{eq:modbb}
        B_\lambda(T_{BB})_m =   B_\lambda(T_{BB}) \left(\frac{\lambda_0}{\lambda}\right)^{\beta},
    \end{equation}
        \noindent where $\beta$ is the power index of the grain emissivity: typically between 0 and 3 for ideal dielectic materials \citep{Helou1989}.   In two of the cases (HIP~74235 and HIP 74926) we have set the excesses to peak at $W2$, since the information from the other \WS\ bands is not sufficient to constrain the temperatures.  For the other two stars we have sought fits that satisfy all of the \WS\ excesses and upper limits. 
        
        HIP 30893 and HIP~109941 are the only stars for which $\beta$ falls between 0 and 3, in agreement with thermal emission from dust with low emissivity.  HIP~74235 and HIP~74926 have grain emissivity indices $\beta>3$ that exceed physical values and are difficult to interpret.   We therefore can not conclude with confidence that dust is at the origin of any of the four K-star $W2$ excesses.
        
	It is possible that the $W2$ excesses from the four K stars are related to their late spectral types, but not for reasons of inaccurate calibration of the photospheric $W1-W2$ color.  Instead, the responsible mechanism may be chromospheric activity.  One of the stars, HIP~109941, is included in the ROSAT Bright Survey catalog \citep{Fischer1998} and possesses H$\alpha$ in emission.  More generally, K stars have relatively active chromospheres compared to earlier-type stars, driven by deep convection.  $W2$ spans the CO fundamental vibration-rotation bands, which are prominent in K stars.  CO could conceivably be observed in emission under the right circumstances.  CO emission at 4.7\um\ is indeed observed in the Sun's lower chromosphere, within 1000~km of the Sun's limb, at gas temperatures of 3000--3500~K \citep{Solanki1994}.  However, the emission does not contribute a significant portion of the Sun's bolometric flux.  K dwarfs are more chromospherically active than the Sun, although it remains to be seen whether their entire $W2$-band fluxes can be raised by 5\%--8\% through CO line emission.

	
	Because the nature of the $W2$ excesses remains speculative, and because the confidence threshold for the detections is lower ($\gtrsim$95\%), we do not count the five stars discussed in this section toward the overall number of debris disks detected in our study.  We single out only the F2V star HIP~96562 as a potential host  of hot (780~K) circumstellar dust.  If this excess is real, it would be among the most tenous debris disks detected around any star.

     \subsection{Circumbinary Dust}\label{sec:BinaryDust}

		The majority of studies looking for IR excesses from circumstellar disk material limit themselves to single stars, as the possibility of photometric confusion or contamination from closely separated stars is a concern. This is also the case in our study, as we aimed to remove all visual binary systems in which a companion may affect the photometry in the different \WS\ bands differently (see \S~\ref{sec:sample_definition}). However, a small number of stars, mostly in very wide binary or multiple systems passed all of our contamination checks and have bona-fide IR excesses.  Only a few close binaries were allowed: those for which the component spectral types were very similar and so the composite $B_T-V_T$ color of the system is representative of the component's colors.

		Using information from the Washington Double Star Catalog\footnote{http://ad.usno.navy.mil/wds/} \citep[WDS;][]{Mason2013} and from the literature, we identified 25 stars from our debris disk candidates that are part of binary or multiple star systems. Projected orbital separations are listed in Table~\ref{tab:binaries}.  Three of these stars have companions projected separations $<12\arcsec$ -- HIP~9141, HIP~16908 and HIP~95261 -- placing them within the $W4$ beam. Thus the flux from these companions might mimic and IR excess attributed to the primary target. However we find this is not the case: HIP~9141 has an equal mass companion \citep{Biller2007} and the SED for this star does not show an excess attributed to a binary component. HIP~95261 has an M7/8 spectral type companion, but the $W4$ flux for this star is $\sim20\%$ above the photosphere and does not possess a significant $W3$ excess. The inferred dust temperature is thus inconsistent with this star's companion. HIP~16908 has an M1/3V companion but the inferred dust temperature, along with the slope of the SED and an insignificant $W3$ excess is inconsistent with the IR flux of an M-type stellar companion. 
        
        We compare the calculated circumstellar dust radius (Table~\ref{tab:diskcandidates}) and the binary separation to infer the location of the dust with respect to the stellar components. 
         
       Most (23) of the projected separations between the stellar binary components are larger than the inferred dust orbital radius, and the dust is therefore circumstellar.  Given sufficiently wide angular separations between the stellar components in most of these systems, we are confident that the debris disk is co-located with the component identified in the \hip\  catalog.  
     
     The remaining 2 stars, HIP~2472 (A0V) and HIP~22394 (K3V) are part of spectroscopic binary systems. There is no information in the literature for the orbital elements or spectral type for the binary component of HIP~2472. The binary component for HIP~22394 has a published orbital period of 11.9 days. The average separation of the stars would $\sim0\farcs1$. The radius for the dust in both these systems is estimated to be at 2.7 and 1.5~AU respectively.  Since our assumption of blackbody dust properties is simplistic, and in reality circumstellar dust grains have poorer emissivity, our inferred dust orbital radii may be too small by a factor of up to two.  Therefore, in these two cases, we conclude that the dust is in circumbinary configuration.

\section{Discussion}
\label{sec:discussion}

\subsection{Comparison to Previous Work}\label{sec:comparison}

	We compare our sample of \hip\ debris disks discovered in \WS\ to those previously reported in published work.  The literature sample consists of excesses detected at multiple reference wavelengths, from IR surveys with \iras, \iso, \spitzer, \akari, \WS, and {\it Herschel} and includes stars not in \hip.  Our compilation of published results contains a total of 449 bona-fide debris disks within 75~pc, most (389) of which satisfy the spatial and color constraints that we placed on our science sample: i.e., $|b| > 5^{\circ}$ and $-0.17 <B_T-V_T \leq 1.40$.  Among these, 261 have known warm component excess emission (10--30\um).   
	
	We have identified 220 debris disks within 75~pc, 108 of which are new detections, and 114 have previously reported mid- and/or far-IR excesses ($\lambda>10$\um).  That is, our study has expanded the overall 75~pc debris disk census by $108/388=28\%$. Ten of the 114 previously known disks were not known to possess excesses at $\lambda<30$\um, so the total number of new 10--30~$\micron$ disk identifications from our study is $108+10=118$: a $118/262=45\%$ increase.  The third column of Table~\ref{tab:excessstats} lists whether our \WS-detected debris disks have previous detections at wavelengths similar to 12~$\micron$ or 22~$\micron$. The Venn diagram in Figure~\ref{fig:Venn}  compares the number of detections in our survey to those stars with IR excesses discovered from past surveys at 10--30\um and at $\lambda \geq 30$~\um.
	     
	Our very strict photometric selection criteria and binarity checks have excluded a significant fraction (33\%) of the overall 75~pc \hip\ sample.  The fact that over half of our 220 debris disk identifications are new indicates that previous searches for debris disks in all-sky surveys are only $\lesssim$50\% complete to the precision limits of \WS.  Hence, there is a potential to further double the number of known warm debris disks outside of the 75~pc \hip\ sample.
	       
     We can also estimate the completeness of our own debris disk identification method by comparing the fraction of \hip\ stars included in our science sample to the fraction of known 10--30$\micron$ debris disks that we recover.  As discussed in \S \ref{sec:science_sample}, our science sample includes 67\% of $|b|>5^{\circ}$ \hip\ 75~pc main sequence stars with $-0.17<B_T-V_T<1.4$.  Within the same constraints we confirm 78\% of the disks known from \WS\ and \akari, and 38\% of the disks known from \spitzer.  We do miss most (14/23) of the few known 10--30$\micron$ debris disks from \iras\ and \iso, only because these stars exceede our $W2>2.8$~mag brightness threshold.
     
     Therefore, our selection is at least as, or more sensitive than any of the previously published work that uses data from all-sky infrared survey telescopes.  We achieve this without compromising confidence in our reported detections, as our overall $W4$ excess selection has 99.5\% reliability.   The reason for the lower fraction of recovered \spitzer\ 10--30$\micron$ excesses is the greater sensitivity of targeted \spitzer\ observations, and the improved ability to remove the stellar photospheric contribution in \spitzer\ IRS observations.  The missed warm excesses known from \spitzer\ are indeed all tenuous, below the sensitivity or precision limits of \WS.
 
    Our search for 5--22$\micron$ excesses from warm debris disks in the solar neighborhood is the most comprehensive and sensitive one to date, with a sample of nearly 8000 stars within 75~pc.  Nevertheless, several recent $W4$-only studies have reported substantial numbers of new debris disk identifications in \WS, with samples that in some cases have significant overlap with ours.  In the following, we compare our findings to these particular ones, and identify areas in which our work represents an improvement.
 
    \subsection{Comparison to the \WS\ $W4$ Debris Disk Study of \citet{Wu2013}}
    \label{sec:comparison2Wu2013} 
 
        \citet{Wu2013} performed a search for $W4$ excesses from bright ($V<10.27$~mag) \hip\ stars, identifying 112 excesses, 70 of which were considered new candidate debris disks. While similar to ours, their analysis differs in ways that make the two studies complementary, with ours being sensitive to  excesses around brighter stars (saturated in \WS), and to altogether fainter excesses around stars within 75~pc. 
        
        \citet{Wu2013} use a sample of 7624 stars within 200~pc, comprised of sources detected at SNR $>20$ in $W4$, parallactic precision better than 10\%, photometric precision better than 2.5\% in $B-V$ colors,  \mass\ $\sigma_{K_s}<0.1$~mag, and unsaturated photometry in $K_s, W3$, and $W4$. Their excess candidates are defined as stars with $K_s-W4$ colors at least $4\sigma$ redward of the mean, where the mean and $\sigma$ are calculated in four bins based on the $J-H$ colors of stars.  This is analogous to our analysis using $B_T-V_T$ rather than $J-H$, and a running mean rather than four bins.  \citet{Wu2013} removed sources contaminated by IR cirrus or confusion after their excess candidates were selected. 
    
	The \citet{Wu2013} approach results in several important differences in the results.  First, \citet{Wu2013} probe stars out to much larger distances than we do, but they confine their analysis to the brightest unsaturated objects, with high-significance $W4$ detections and precise optical photometry.  This allows the detection of disks with low fractional luminosity around any star in their sample, but at the same time rejects both the brightest saturated stars and fainter stars with G or K spectral types, around which we have detected significant excesses. If we compare the $W4$-excess disks in our science sample ($<75$~pc) to their selection criteria, we find that $180/220=82\%$ of our science sample disks are removed from their study: mostly because of saturation in $K_s$ or because their $B-V$ color errors are $>2.5\%$.

	Second, \citet{Wu2013} choose to eliminate some sources of contamination \textit{after} performing their color selection.  On the one hand, this allows them to retain a larger statistical sample of stars to characterize the full $K_s-W4$ distribution. On the other hand, it results in a higher probability of missing faint excesses: including stars with \WS\ photometry contaminated by line-of-sight IR cirrus systematically increases the width of the $K_s-W4$ distribution.  Our stricter selection criteria result in a cleaner sample, with $Wi-Wj$ distribution widths almost entirely accounted for by the photometric uncertainties (\S~\ref{sec:IRAnalysis}).

	   Our use of \WS-only colors and our treatment of the photometric systematics (\S\S\ref{sec:atlas_vs_single}--\ref{sec:IRAnalysis}) also allows us to potentially detect fainter excesses. \citet{Wu2013} use \mass\ $K_s$ photometry where the observations were conducted years prior to the launch of \WS. \mass\ minus \WS photometry is vulnerable to precision limitations induced by stellar variability or cross-platform systematics.  These also increase the width of the $K_s-W4$ color distribution and can result in missed excesses. %
	
	Finally, we note that the tenuous excesses reported in \citet{Wu2013} from six F stars within 75~pc--- HIP~22531, HIP~29888, HIP~42753, HIP~67953, HIP~70386, and HIP~72138---are likely not caused by circumstellar dust, but are the result of the stars' known binary companions.  \citet{Wu2013} do note the presence of known companions in all of these cases, although do not rule out debris disks.  We observe that the $K_s-W4$ excesses for these stars are similar to their respective $K_s-W1$, $K_s-W2$, and $K_s-W3$ excesses.  In most of these cases the wider \WS\ beam has not  resolved close visual binaries that are otherwise partially resolved in the seeing-limited \mass\ observations.  In the case of the eclipsing binary HIP~72138 the \mass\ and \WS\ osbervations have likely seen the system at different orbital phases, such that the measurements are discrepant and a small excess appears to exist at \WS\ wavelengths.
	
	While we do not address M stars in our study, we also note that two of the three M stars within 75~pc, HIP~21765 and HIP~63942, identified as candidate debris disk hosts in \citet{Wu2013} are also close ($1\farcs4$--$2\farcs0$) visual binaries.  These are partially resolved in \mass\ and their $K_s-W4$ excesses are similar to those at the rest of the $K_s-\WS$ colors.  That is, the excesses are most likely not from dust.
	
     We do not recover every single reported debris disk in \citet{Wu2013}.   Within 75~pc we recover 37 of the 47 bona-fide debris disks reported in \citet{Wu2013}, where we have excluded the eight F- and M-star binaries discussed above.  The remaining 10 stars did not pass our selection criteria (\S~\ref{sec:IRAnalysis}), designed to remove objects for which the photospheric calibration of \WS\ colors is uncertain, and which may produce false-positive detections.  HIP~12351 is an M star, excluded by our $B_T-V_T<1.4$~mag criterion.  HIP~11360 has contaminated \WS\ photometry (\WS\ confusion flag set to `dddd', indicative of contamination from a diffraction spike in each band by a closely separated star\footnote{\url{http://wise2.ipac.caltech.edu/docs/release/allsky/expsup/sec2\_2a.html}}), although the $W4$ excess does appear real.  HIP~20713 has a companion within 5$\arcsec$ listed in the \hip\ Visual Double Database. Lastly, seven of the stars within 75~pc in \citet{Wu2013} are giants (HIP~12361, HIP~15039, HIP~26309, HIP~43970, HIP~53824, HIP~55700, and HIP~100787), whereas we have focused only on main sequence stars.

	Altogether, because of the greater emphasis on uncontaminated photometry, our analysis has resulted in greater sensitivity to debris disks and a larger detection rate within 75~pc.  We have missed only one of the bona-fide main sequence B--K star debris disks from \citet{Wu2013}---HIP~11360, excluded because of contamination flagging in \WS.  That is, we are 100\% complete to debris disks within our overall set of constraints.  Conversely, the \citet{Wu2013} study encompasses a larger volume and identifies more distant debris disk systems.  However, it does not include stars brighter than the $K_s\approx4.2$~mag saturation limit in \mass, whereas we are able to.  In addition, extra scrutiny is required to remove spurious excess identifications associated with double star systems.

\subsection{Comparison to \WS\ $W4$ Debris Disk Study of \citet{Cruz-SaenzdeMiera2013}}
\label{sec:comparison2Cruz2013}

    \citet[][henceforth CS14]{Cruz-SaenzdeMiera2013} also carried out a search to find $W4$ excesses around main-sequence stars, finding 197 disk candidates.    Their method to search for excesses is similar to ours, in that they relied solely on \WS\ photometry (the $W2-W4$ color) to identify excesses while avoiding external systematics and stellar variability.  CS14 focused on unsaturated F2-K0 stars with $V<15$mag that were free of contamination in \WS.

	Because of the elimination of saturated stars in CS14 and our focus on stars within 75~pc, the two studies are almost entirely complementary.  In particular,  there is no overlap in the reported detections. This is because their parent sample is generated from SIMBAD, and most of their stars are not in the \hip\ database: only 68 of their 197 disk-host stars have \hip\ parallaxes.  Only 3 of these are within $<$75~pc.  We confirm two of these: HIP~5462 and HIP~93412.  The remaining star, HIP~63880, is within 5$^{\circ}$ of the galactic plane, and so is not included in our selection, although the excess reported in CS14 is likely real. 
  
\subsection{Comparison to Vican \& Schneider (2014) }    

Recently, a study of the age dependence of $W4$ excesses was published by \citet{Vican2014}.  In a sample of 2820 \hip\ field FGK stars with ages estimated from chromospheric activity, \citet{Vican2014} report 98 excesses, 74 of which are identified as new, for a detection rate of 3.5\%.  The authors use photospheric fitting of the stellar SED, from the $BVJHK_s$ photometry, which they then compare to the measured $W4$ flux and error.  The quality of the photospheric fits is inspected visually, and in the absence of nearby contamination evident from the \WS\ images, excesses with SNR $>5$ are deemed significant.

Eighty-one of the 98 excesses reported in \citet{Vican2014} are from stars within 75~pc from the Sun, and would therefore be expected to be within our science sample, modulo the set of constraints that we impose to retain stars with clean \WS\ photometry.  Among these we recover 24 of the reported excesses, we miss 11 stars because of our selection criteria, and do not confirm the remaining 46 excesses, even though those stars are included in our analysis.

We find that the 46 unconfirmed excesses from \citet{Vican2014} have \ES\ values that are often well below the 99.5\% confidence threshold in our $W1-W4$, $W2-W4$, and $W3-W4$ color distributions.  A select few are even negative: e.g., HIP~117247, identified as a 6$\sigma$ $W4$ excess in \citet{Vican2014}, or HIP~10977, which has a negative $\Sigma_{E[W1-W4]}$ and $\Sigma_{E[W2-W4]}$ along with a positive but insigificant $\Sigma_{E[W3-W4]} = 0.49$. 

We believe that our empirically determined 99.5\% confidence threshold in $W4$ is robust, and is as aggressive as the data allow: evidenced by our 100\% recovery rate of B--K main sequence star debris disks within 75~pc reported in \citet{Wu2013}.  Conversely, it is likely that the excess selection technique employed by \citet{Vican2014} is subject to unrecognized stellar variability between the multiple epochs that span the collection of the $BVJHK_s$ and \WS\ photometry.  The fitting of stellar photospheres from the $BVJHK_s$ photometry, independently of any of the \WS\ measurements, and the subsequent selection of $W4$ excesses above the fitted photosphere, biases the excess candidate selection toward stars that are overall slightly brighter during the \WS\ epoch.  In addition, such an approach should incorporate the overall 1.5\% uncertainty in the \WS\ $W4$ calibration \citep{Wright2010}.  Our empirical calibration of the stellar photospheric colors in \WS\ and our use of \WS-only photometry for excess selection allows us to calibrate both of these sources of systematic error.

Overall, we find that the 10--30$\micron$ excess rate for field FGK stars in the \citet{Vican2014} study is approximately 1/3 of their reported one, and so more in agreement with the rate that we estimate in \S\ref{sec:disk_fraction} below.

    \subsection{Stellar Spectral Type and Warm Disk Fraction}
    \label{sec:disk_fraction}

    As detailed in \S\S\ref{sec:comparison}--\ref{sec:comparison2Cruz2013}, because of our strict selection criteria, our study is not complete to all warm debris disks around \hip\ stars within 75~pc.  Nonetheless, within our carefully selected and unbiased science sample, we have performed the most sensitive and complete photometric identification of 10--30$\micron$ excesses around main sequence stars using \WS.  In the following, we use this result to study the relative occurrence of warm debris disks in the solar neighborhood.
   
       Figure~\ref{fig:wise_vs_all_spt} plots the distribution of detected 10--30$\micron$ excesses from \WS\ and previous surveys as a function of spectral type, within the spatial and color constraints of our science sample.  We find that \WS\ detects approximately five times as many warm debris disks as \iras\ and \akari\ combined.  Our particular study also increases by 45\% the number of known warm dust excesses within 75~pc.  Notably, we detect a substantial number of disks around cool stars, where the disks are intrinsically fainter.  The discovery of these fainter disks is a consequence of both the increased sensitivity of \WS\ compared to \iras\ and \akari, and of our careful calibration of \WS\ systematics.
        
        We present the distribution of \WS\ excess occurence rate as a function of stellar $B_T-V_T$ color and spectral type in Figure~\ref{fig:wise_BV_w3w4}.  We find that B8--A9 stars show a $21.6$\%$ \pm 2.5$\% incidence of significant $W4$ excesses, and a $1.0$\%$ \pm 0.5$\% incidence of $W3$ excesses.  Solar-type FGK stars have much lower excess occurrence rates: $1.8$\%$ \pm 0.2$\% at $W4$ and $0.08$\%$ \pm 0.04$\% at $W3$.  The occurrence rates represent the results for the most sensitive among the different color combinations.
        
        Our findings are in broad agreement with previous searches for $W4$ excesses on \WS, although we have had to point out several caveats with previous such studies.  Thus, \citet{Wu2013} report that 6.9\% of main-sequence FGK stars possess $W4$ excesses detected at the 3$\sigma$ level.  However, without detailed attention to photometric systematics they have adopted a higher working threshold for excess detection---4$\sigma$--at which level only 2.2\% of their FGK stars have $W4$ excesses.  We also discussed that a fraction ($\approx$25\%; \S\ref{sec:comparison2Wu2013}) of the excesses identified in \citet{Wu2013} do not originate from dust, or are not associated with main sequence stars.  That is, the actual rate of identifications of main sequence debris disks in \citet{Wu2013} is $\approx$1.6\%.   Similarly, CS14 report that 2\% of all their FKG main-sequence stars possess 3$\sigma$ $W4$ excesses, while our correction to the FGK debris disk rate found in \citet{Vican2014} is 1.2\%.  All of these warm-disk ($\lesssim$150~K) excess rates are consistent with our own findings for incidence of $W4$ excesses around FGK stars.  
     
        Compared to previous unbiased studies of warm debris disks with \spitzer\, our \WS\ analysis produces a factor of 1.5--3 lower detection rates.  \citet{Su2006} determine a 32\% rate of debris disks among A stars at 24$\micron$, while \citet{Carpenter2009} find debris disks with 10--70$\micron$ excesses around 3\% of $>$300 Myr old FGK stars.  The discrepancies with the \spitzer\ studies are attributable to the higher sensitivity of pointed \spitzer\ observations.  

	Finally, our 0.08\%--1.0\% 12$\micron$ excess rate from exozodi ($\sim$300~K) among field stars is in agreement with an estimate from \WS\ in  \citet[0.01\%;][]{Kennedy2013} and with findings from \spitzer\ \citep[1\%;][]{Lawler2009}.  We note that our calibration and sample selection approach have enabled a somewhat better sensitivity to exozodi than the previous \WS\ study.  In addition, our large-scale study has now for the first time provided a sufficient sample size to establish the relative frequency of exozodi between A and FGK stars: bright ($f_d>10^{-4}$) exozodi are a factor of $\sim$10 more common around hot stars than around solar analogs.

\section{Conclusion}\label{sec:conclusion}
        
         We identify a volume-limited sample of \hip\ stars within 75 pc that show infrared excess fluxes based on photometry contained in the \WS\ All-Sky Data Release. We carefully screen the \WS\ photometry for various sources of false-positives both astrophysical and instrumental. One such issue, newly identified in our work, is that in a tiny fraction of \WS\ photometry, the median of single-exposure fluxes is inconsistent with the \WS\ All Sky Catalog flux, and neither is reliable.  We reject photometry compromised by this and other issues; precisely calibrate flux-dependent systematic effects in saturated photometry; and correct for the dependence of \WS\ colors on photospheric temperature. Using the blue wing of the resulting color distributions to empirically evaluate our false positive rate (FPR) for the red outliers that correspond to dusty circumstellar disks, we robustly detect 215 such disks at 22$\micron$ with FPR $<$ 0.5\% and 5 additional disks at 12$\micron$ with FPR $<$ 2\%.

Our careful screening and precise calibration of the \WS\ photometry enables us to identify faint circumstellar dust disks that had gone unnoticed in previous analyses, in addition to confirming disks that had been previously detected using photometry from \WS\ and other missions.  Our new detections represent, in total, an increase of 45\% in the number of stars within 75~pc known to have flux excesses at mid-IR wavelengths. In contrast to {\it IRAS} and {\it ISO}, which produced many detections of cold circumstellar dust, the \WS\ mid-infrared bands have enhanced sensitivity to warmer dust in regions analogous to our own Solar System's asteroid belt and zodiacal cloud --- regions most likely responsible for terrestrial planet formation. We report the following detections:

		\begin{enumerate}
        \item 220 stars with FPR$<$0.5\% mid-IR excesses at 22$\micron$ and/or FPR$<$2\% excesses at 12$\micron$. For 113 of these we present the first detection of a debris disk at any wavelength, and for a futher 10 that have known longer-wavelength excesses, we present the first measurement of an excess at $12\micron$ and/or $22\micron$.
        
        \item A subset of 211 of our disks are detected with significant excesses in 22$\micron$ only. Aggregate 12$\micron$ excesses can be detected by weighted averages of the 12/22$\micron$ excess flux ratio over different subsets of this sample, and these aggregate 12$\micron$ detections are highly significant.  The subset with previously published low (50--120 K) dust temperatures has an aggregate 12/22$\micron$ excess flux ratio consistent with low-temperature dust, while the aggregate flux ratio for the previously unknown disks indicates that many of them have dust at asteroidal temperatures ($>$ 130 K).

        \item A subset of 4 stars possess significant excess detections at both 12 and 22$\micron$, with a flux ratio indicative of dust temperatures ranging from $\sim 130$~K to $\sim 200$~K. All of these systems are known to possess long-wavelength ($>60\mu m$) excesses well fit by colder dust, and none were suspected to have $>$100~K dust.  Hence, our results indicate the presence of dust at multiple temperatures in these systems.

        \item A subset of 5 disks are detected with signficant excesses only at 12$\micron$.  Upper limits to the 22$\micron$ excesses in these systems yield 3$\sigma$ lower limits on the temperature ranging from $\sim 175$~K to $\sim 275$~K.  While the coolest of these limits would permit asteroidal-temperature dust, the data are more consistent with warmer dust.  Such dust would overlap with the habitable zones in these systems and could come from planetesimals left over from the formation of terrestrial planets.

        \item Five additional stars, not included in our count of 220 detected dust disks, possess shorter-wavelength excesses at 4.6$\micron$ with FPR$<$ 5\%.  One of these excesses, around the F2V star HIP~96562, is suspected to be caused by hot (780~K) dust.  The origin of the remaining four excesses, all associated with K dwarfs, remains speculative.  It is possible that in two of the cases the thermal emission is caused by tenuous amounts of hot, short-lived, sub-micron-sized dust.  However, this scenario can not account for all four cases of $W2$ emission from K stars.  We therefore suggest an alternate explanation involving chromospheric activity.
                        
        \item $1.8$\%$ \pm 0.2$\% of solar type (FGK) stars and $21.6$\%$ \pm 2.5$\% of A stars possess mid-infrared excesses at 22$\micron$, and the median lower limit to the fractional dust luminosity is $L_{\rm dust}/L_\ast\gtrsim 1.2\times10^{-6}$ for the A stars. At 12$\micron$, the occurrence rate of excesses is $0.08$\%$ \pm 0.04$\% for solar type stars and $1.0$\%$ \pm 0.5$\% for A stars.

          \item As a result of our study, the number of debris disks with known 10--30$\micron$ excesses within 75~pc (379) has now surpassed the number of disks with known $>$30$\micron$ excesses (289, with 171 in common), even if the latter are known to have a higher occurrence rate in unbiased samples.

        \end{enumerate}

\begin{figure}
\centering
\includegraphics[scale=0.45]{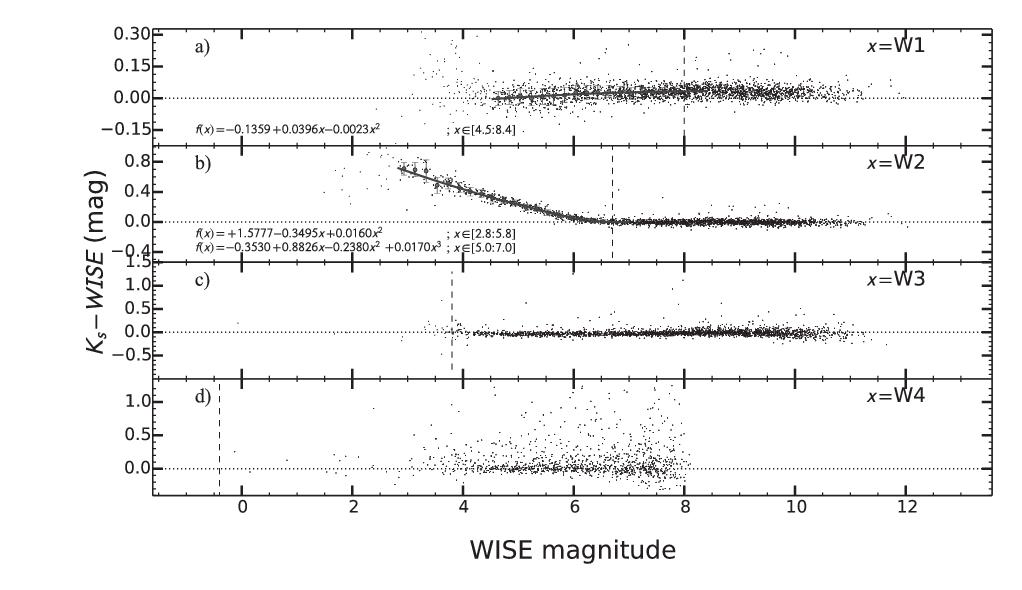}
\placefigure{figure2}
\caption{\mass\ $K_s$-\WS\ vs.\ \WS\ relations used for correcting systematics in  saturated $W1$ ({\bf a}) and $W2$ ({\bf b}) photometry.  The empirical $K_s$-\WS\ vs.\ \WS\ distributions are a combination of bright B8--A9 dwarf stars from our science sample with fainter $B-V<0.10$~mag A0 stars from the Tycho-2 Spectral Type Catalog \citep{Wright2003}. Saturation limits for each \WS\ band are shown with vertical dashed lines. Polynomials were fit to the saturated portions of the $K_s-W1$ vs.\ $W1$ and $K_s-W2$ vs.\ and $W2$ distributions to model the systematic trends and correct the saturation. Two polynomials were fit to the saturated $W2$ data to account for the knee between 5.4~mag and 6.7~mag.  A discontinuity of $\lesssim$0.010~mag was allowed at $W2=5.4$~mag. The $W3$ ({\bf c}) and the $W4$ ({\bf d}) photometry appears self-consistent throughout and does not require correction.     
}
 
\label{fig:wise_flux_biases}
\end{figure}

\begin{figure}
\centering
\begin{tabular}{cc}
\includegraphics[scale=0.4]{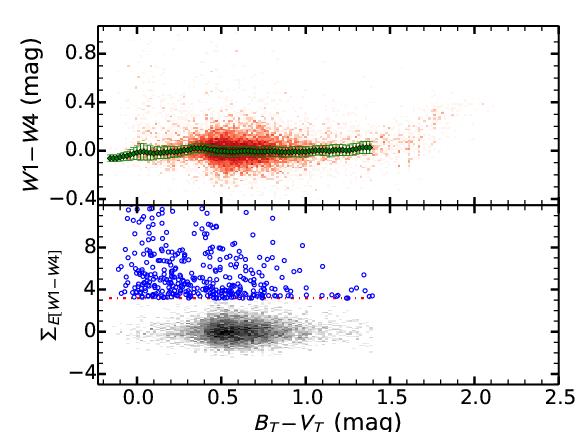}&
\includegraphics[scale=0.4]{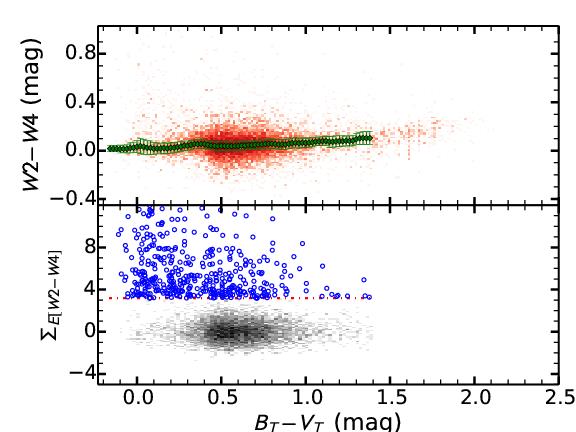}\\
\includegraphics[scale=0.4]{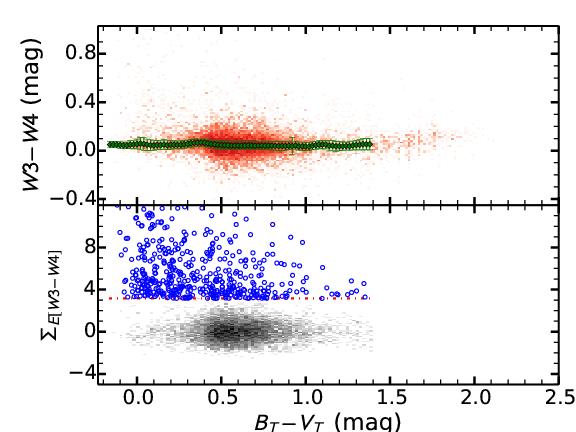}& 
\includegraphics[scale=0.4]{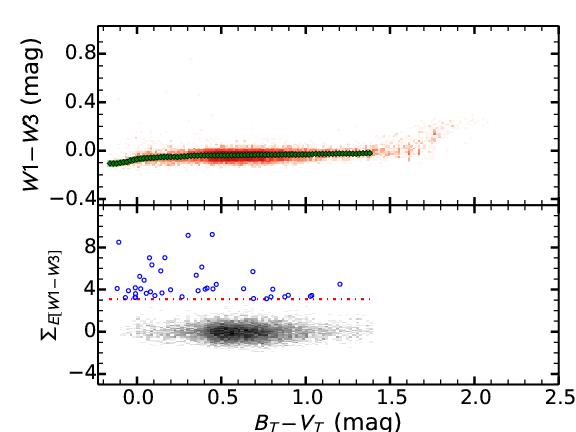}\\
\includegraphics[scale=0.4]{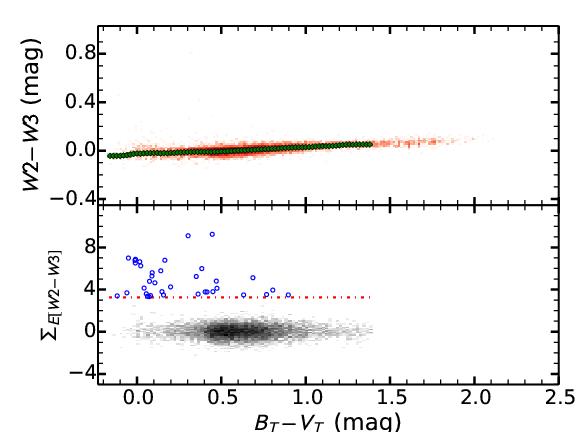}&
\includegraphics[scale=0.4]{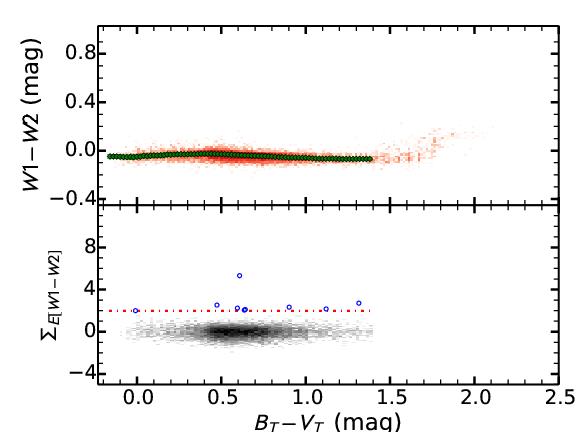}

\end{tabular}
\caption{\textbf{Top Half of Each Panel}: \WS\ vs.\ Tycho-2 $B_T-V_T$ color-color diagrams of our parent sample stars (red). The green diamonds in each panel follow the running mean of the parent sample. 
We eliminated stars outside the $-0.17<B_T-V_T<1.4$~mag range from all of our analysis. \textbf{Bottom Half of Each Panel}: Plots of the significance \ES\ of the color excess as a function of $B_T-V_T$. These are residuals of the subtraction of the photospheric running mean, normalized to the 1~$\sigma$ scatter. The stars selected as debris disk candidates in the parent sample are denoted by open blue circles. These are more significant than the confidence limit $CL$ thresholds shown by the dashed purple lines.  $CL=$99.5\% for $W4$ excesses, 98\% for $W3$ excesses, and 95\% for $W2$ excesses.}
\label{fig:Res}
\end{figure}

\begin{figure}
\centering
 \begin{tabular}{cc}
 \includegraphics[scale=0.4]{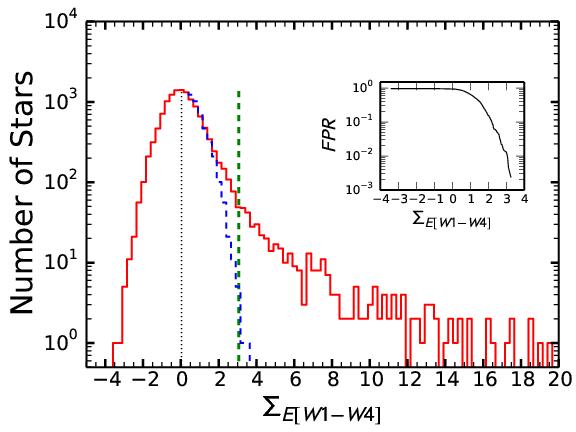}&
 \includegraphics[scale=0.4]{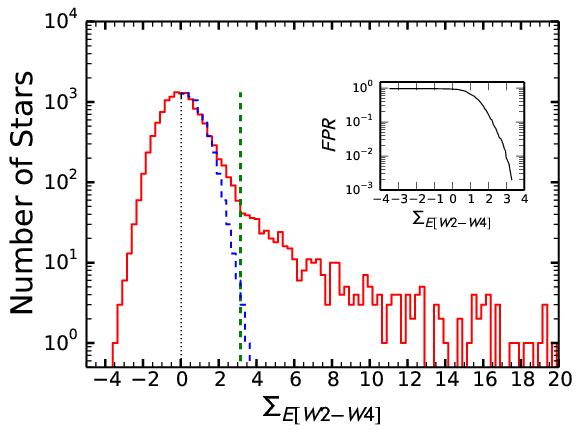}\\
 \includegraphics[scale=0.4]{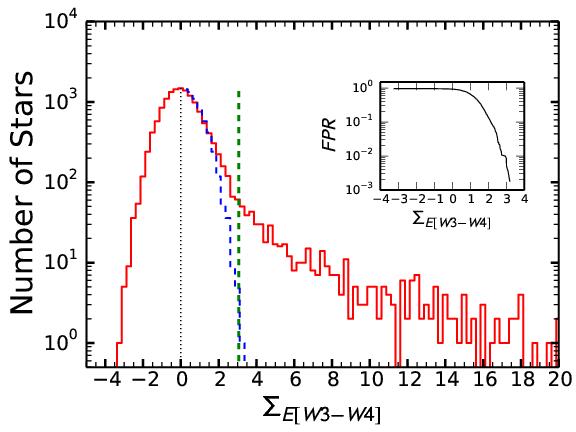}&
 \includegraphics[scale=0.4]{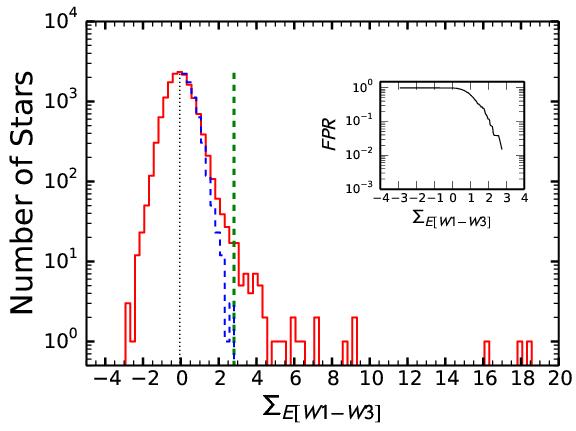}\\
 \includegraphics[scale=0.4]{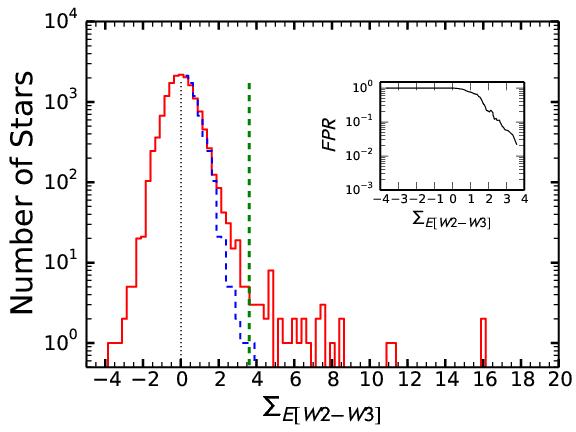}& 
 \includegraphics[scale=0.4]{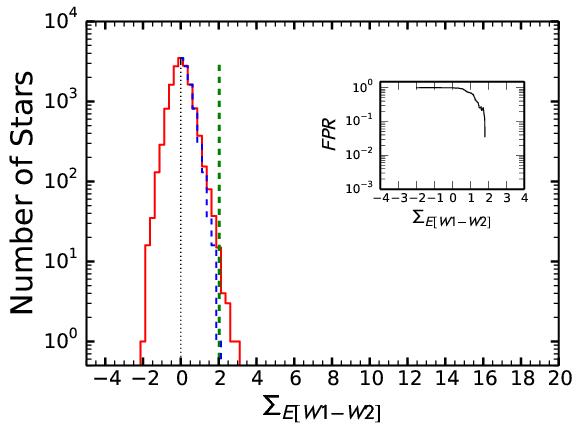}
 \end{tabular}
\caption{Distributions of the significance of the color excess \ES for the stars in our parent sample for each \WS\  color. We assume that the negative excesses, where \ES~$<$~0, are representative of the intrinsic random and systematic noise in the data. A reflection of the negative excess histogram around 0 (dashed histogram) is thus representative of the false positive excess expectation. We define the FPR at a given \ES as the ratio of the cumulative numbers of $>$\ES excesses in the positive tails of the dashed and solid histograms.  The vertical dashed lines indicate the FPR thresholds for each $Wi-Wj$ color, above which we identify all stars as probable debris disk hosts. The insets show the FPR for each distribution.}
\label{fig:colordist}
\end{figure}

\begin{figure}
\centering
\includegraphics[scale=0.8]{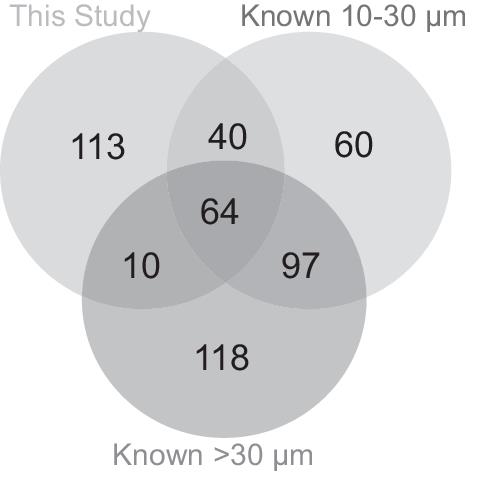}
\caption{A comparison between excess detections in this study and previously reported excesses at mid-IR (10--30\um) and far-IR ($\lambda \geq 30$~\um) wavelengths. Only stars that are within 75~pc of the Sun with galactic latitudes $5^{\circ}$ above or below the galactic plane are included in this comparison.  Our study is focused on \hip\ stars, while the previous studies include non-\hip\ stars, too.  Data for these stars was obtained from the following sources: \citet{Sylvester2000}, 
\citet{Habing2001}, 
\citet{Metchev2004}, 
\citet{Beichman2005}, 
\citet{Chen2005a},
\citet{Chen2005b}, 
\citet{Low2005}, 
\citet{Beichman2006}, 
\citet{Beichman2006a}, 
\citet{Chen2006}, 
\citet{Moor2006}, 
\citet{Smith2006}, 
\citet{Su2006}, 
\citet{Rhee2007}, 
\citet{Rhee2007a}, 
\citet{Trilling2007}, 
\citet{Wyatt2007}, 
\citet{Hillenbrand2008}, 
\citet{Meyer2008}, 
\citet{Rebull2008}, 
\citet{Rhee2008}, 
\citet{Roberge2008}, 
\citet{Trilling2008}, 
\citet{Bryden2009}, 
\citet{Carpenter2009}, 
\citet{Dahm2009}, 
\citet{Kospal2009}, 
\citet{Lawler2009}, 
\citet{Moor2009}, 
\citet{Morales2009}, 
\citet{Plavchan2009}, 
\citet{Su2009}, 
\citet{Koerner2010}, 
\citet{Moerchen2010}, 
\citet{Smith2010}, 
\citet{Dodson-Robinson2011}, 
\citet{Eiroa2011}, 
\citet{Moor2011}, 
\citet{Morales2011}, 
\citet{Zuckerman2011}, 
\citet{Kennedy2012},
\citet{Urban2012} and \citet{Wu2013}.}
\label{fig:Venn}
\end{figure}

	    In addition to the scientific results, notable numerical and tabular references from the present study include:
        
        \begin{enumerate}
        
	    \item the determination of photospherice \WS\ colors from $-0.15 < B_T-V_T < 1.4$~mag main sequecne stars (Table~\ref{tab:wise_bv_trends})
		
		\item polynomial relations for correcting saturated \WS\ $4.5<W1<8.4$~mag and $2.8<W2<7.0$~mag photometry (Figure~\ref{fig:wise_flux_biases})

		\item corrected $W1$ and $W2$ photometry for saturated \hip\ stars with $4.5<W1<8.4$~mag and $2.8<W2<7.0$~mag (Table~\ref{tab:stellarparameters})

        \end{enumerate}
    
         \WS\ has rekindled the search for new disk bearing stars due to its enhanced resolving power compared to previous all-sky surveys like {\it IRAS}, combined with its wider coverage relative to pointed surveys using \spitzer.  Although \WS\ cannot detect disks as faint as \spitzer, for that very reason the brighter, \WS-selected systems are excellent targets for resolved imaging observations, e.g., with the Gemini Planet Imager, ALMA, the LBTI nuller, or the JWST. Such observations would further constrain the structure of the disks and the properties of the dust grains that reside in them, expanding our knowledge of the range of planetary system architectures in the galaxy.

\acknowledgments This publication makes use of data products from the Wide-field Infrared Survey Explorer, which is a joint project of the University of California, Los Angeles, and the Jet Propulsion Laboratory/California Institute of Technology, funded by the National Aeronautics and Space Administration. We also use data products from the Two Micron All Sky Survey, which is a joint project of the University of Massachusetts and the Infrared Processing and Analysis Center/California Institute of Technology, funded by the National Aeronautics and Space Administration and the National Science Foundation. This research has also made use of the SIMBAD database, operated at CDS, Strasbourg, France. This research has made use of the Washington Double Star Catalog maintained at the U.S. Naval Observatory. We would also like to thank Kendra Kellogg for her help in visually inspecting the \WS\ images in the initial stages of this study as well as Joe Trollo for his help in the development phase of the SED plotting algorithm. Most of the figures in this work were created using Matplotlib, a Python graphics environment \citep{Hunter2007}. This work is partially supported by NASA Origins of Solar Systems through subcontract No. 1467483. 

\clearpage

\bibliography{DebrisDisks}

\clearpage

\input{table1.tex}

\begin{deluxetable*}{lccccc}
    \tablewidth{0pt}
    \tabletypesize{\scriptsize}
    \tablecaption{\WS\ IR Excess Selection Summary\label{tab:color_analysis_summary}}
    \tablehead{\colhead{Color} & \colhead{$\Sigma_{E_{\rm CL}}$} & \colhead{Stars in} & \colhead{Stars in} & \colhead{Excesses in} & \colhead{Final Disk Candidates} \\
                \colhead{} &\colhead{} &\colhead{Parent Sample($<$120~pc)} &\colhead{Science sample ($<$75~pc)} &\colhead{ Science Sample ($<$75~pc)} &\colhead{($<$75~pc)} 
                }
    \startdata
	$W1-W4$ &  3.19  &  12942 & 6294 & 133 & 121 \\
    $W2-W4$ &  3.26  &  13203 & 6507 & 164 & 155 \\
    $W3-W4$ &  3.16  &  14434 & 7198 & 208 & 198 \\    
    $W1-W3$ &  2.82  &  15017 & 6788 &   9 & 8 \\ 
    $W2-W3$ &  3.70  &  15245 & 6962 &   4 & 4 \\
    $W1-W2$ &  2.03  &  15053 & 6804 &   8 & 6\tablenotemark{a} \\    
    \hline\hline\\
    Total   & \nodata  & 16960  & 7937 & 243 & 220
    \enddata
\tablecomments{Summary of the results from our \WS\ excess and debris disk candidate identification. $\Sigma_{E_{CL}}$ is the confidence level $CL$ threshold adopted for any given color SNR.  $CL=99.5\%$ for $W4$ excesses, 98\% for $W3$ excesses, and 95\% for $W2$ excesses (\S~\ref{sec:IRAnalysis}). The number of stars in the parent and science samples are those that pass the selection criteria in $\S$~\ref{sec:sample_definition}. The excesses in the science sample are for stars that pass the corresponding excess selection criteria at confidence $\geq$CL.
 The final debris disk candidates are the subset of excesses that survive visual inspection. Rejected sources are listed in Table~\ref{tab:rejects}. The last row lists the total number of unique stars in each applicable column.}
\tablenotetext{a}{The six stars with detected $W1-W2$ excesses are not included in the total number of disk candidates in this study, as described in \S \ref{sec:W2Excess}.}
\end{deluxetable*}

\clearpage
\LongTables
\input{table3.tex}

\clearpage
\LongTables
\input{table4.tex}

\clearpage

\LongTables
\input{table5.tex}

\clearpage


\input{table6.tex}
\clearpage

    
\input{table7.tex}
\clearpage

\input{table8.tex}
\clearpage


\clearpage
\begin{figure}
\centering
\includegraphics[scale=0.7]{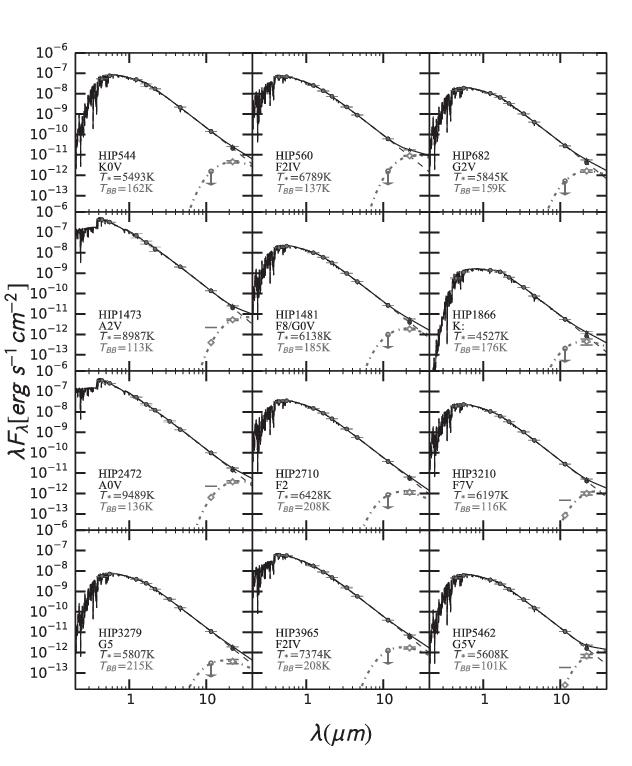}
\caption{SEDs of probable debris disk-host stars in our science sample. The dashed  lines and solid data points correspond to the fitted model NextGen photosphere and to $BVJHK_s$ photometry from the \hip\  Catalogue and \mass\  Point Source Catalog. Fluxes plotted as closed circles were used in the fit, and fluxes plotted as stars---excesses above the photosphere---were not used in the fit.  Cool blackbody curves (dash-dotted line) were fitted to the excess fluxes (open diamonds) at the $W3$ and/or $W4$ wavelengths. The combined photosphere and excess emission for each star is plotted as a solid black line.}
\label{fig:SED}
\end{figure}
\clearpage

\begin{figure*}
\centering
\includegraphics[scale=0.7]{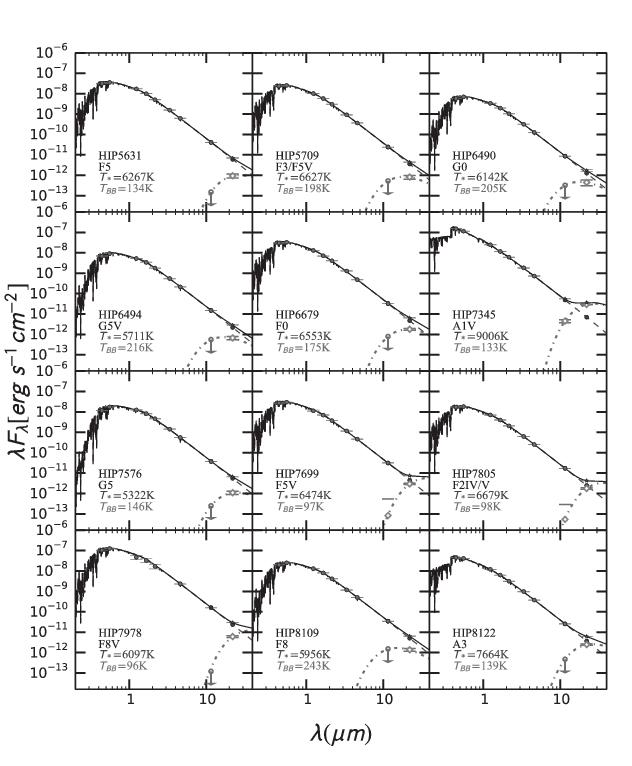}
\setcounter{figure}{5}
\caption{continued.}
\end{figure*}
\clearpage

\setcounter{figure}{5}
\begin{figure*}
\centering
\includegraphics[scale=0.7]{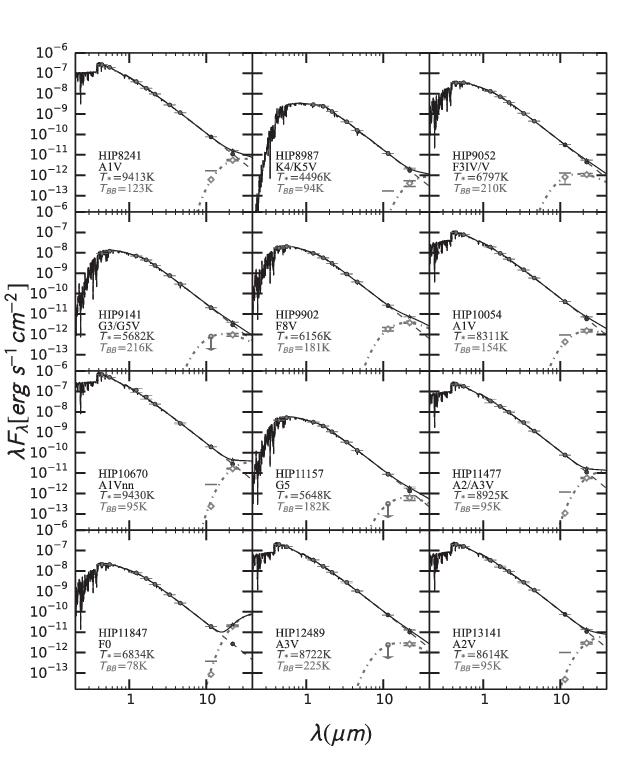}
\caption{continued.}
\end{figure*}
\clearpage

\setcounter{figure}{5}
\begin{figure*}
\centering
\includegraphics[scale=0.7]{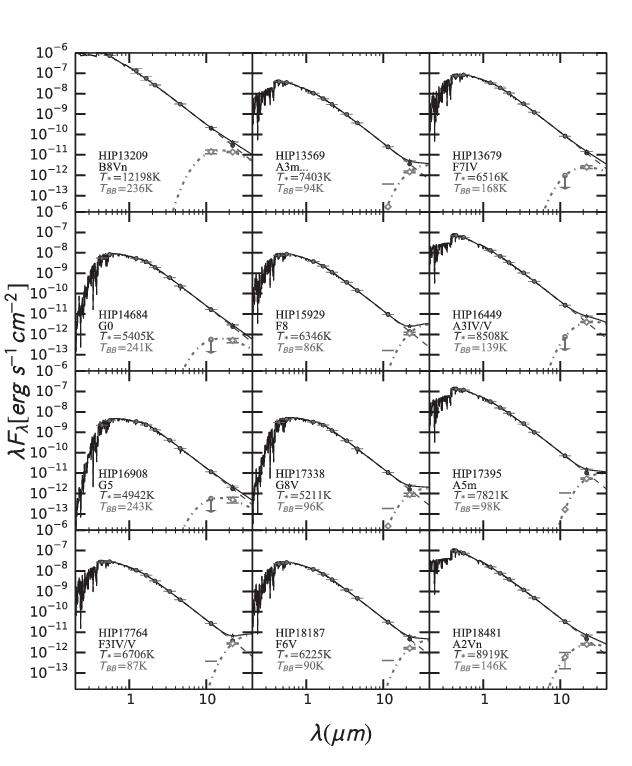}
\caption{continued.}
\end{figure*}
\clearpage

\setcounter{figure}{5}
\begin{figure*}
\centering
\includegraphics[scale=0.7]{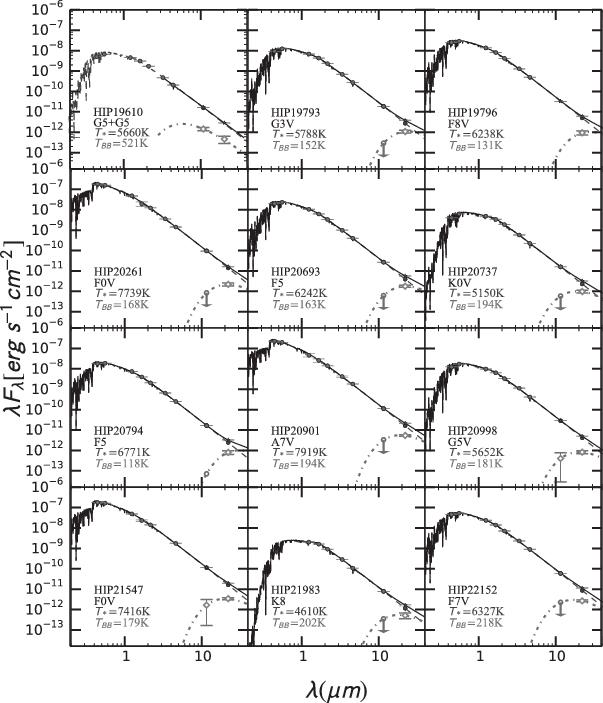}
\caption{continued.}
\end{figure*}
\clearpage

\setcounter{figure}{5}
\begin{figure*}
\centering
\includegraphics[scale=0.7]{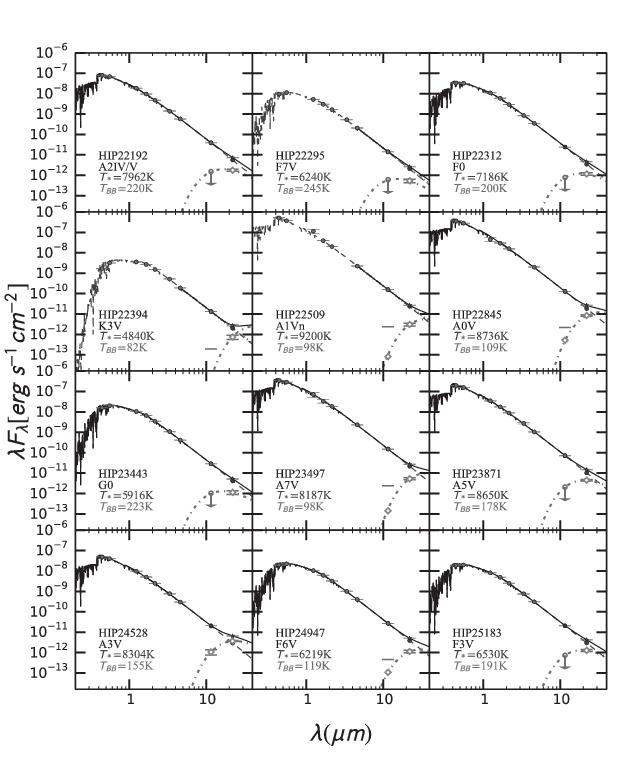}
\caption{continued.}
\end{figure*}
\clearpage

\setcounter{figure}{5}
\begin{figure*}
\centering
\includegraphics[scale=0.7]{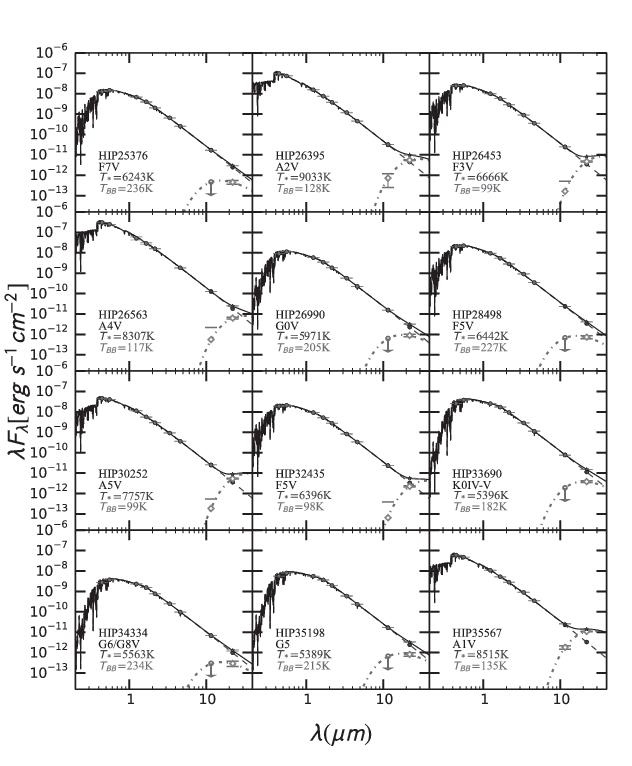}
\caption{continued.}
\end{figure*}
\clearpage

\setcounter{figure}{5}
\begin{figure*}
\centering
\includegraphics[scale=0.7]{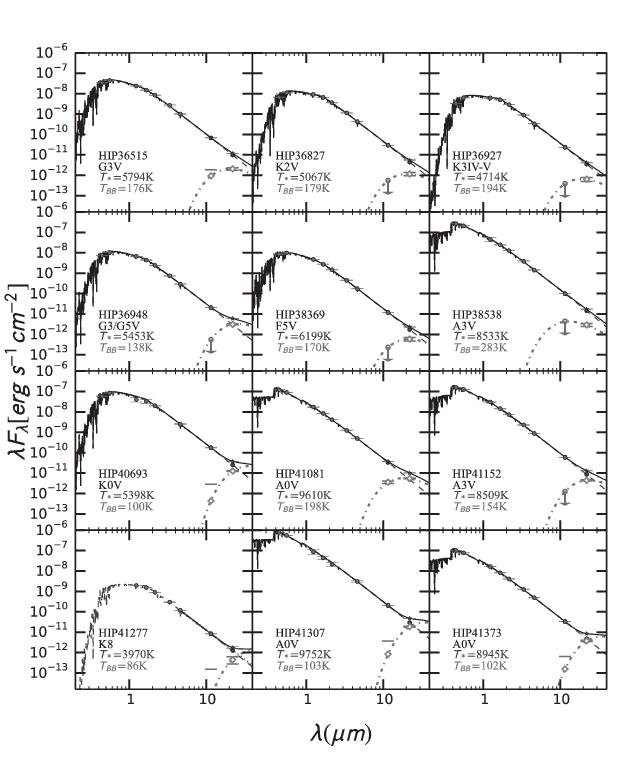}
\caption{continued.}
\end{figure*}
\clearpage

\setcounter{figure}{5}
\begin{figure*}
\centering
\includegraphics[scale=0.7]{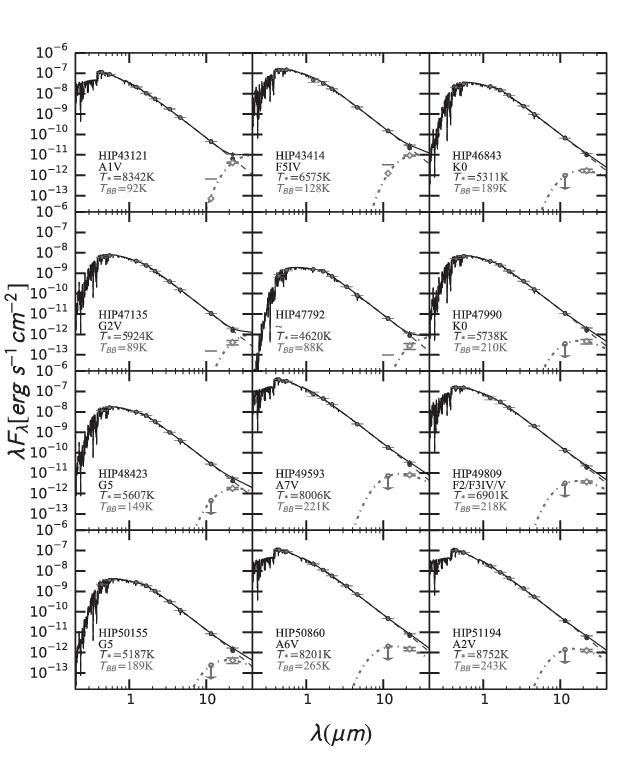}
\caption{continued.}
\end{figure*}
\clearpage

\setcounter{figure}{5}
\begin{figure*}
\centering
\includegraphics[scale=0.7]{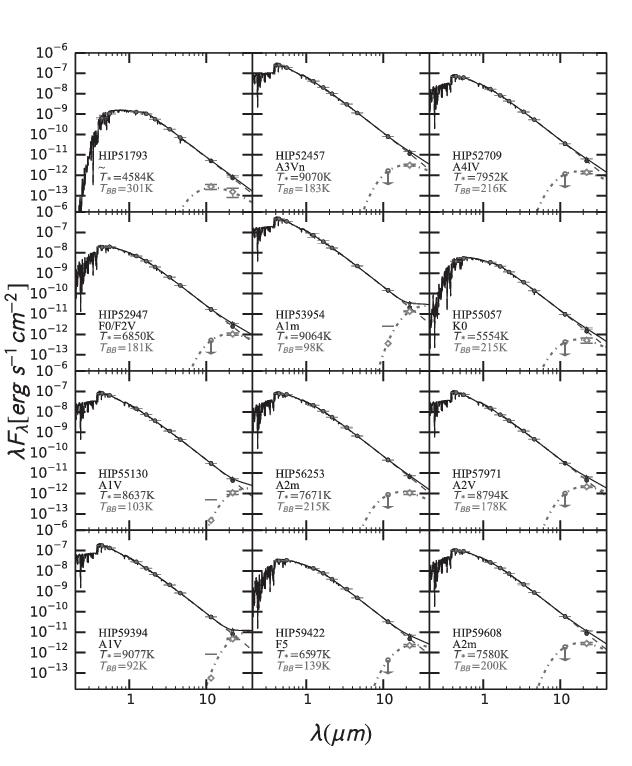}
\caption{continued.}
\end{figure*}
\clearpage

\setcounter{figure}{5}
\begin{figure*}
\centering
\includegraphics[scale=0.7]{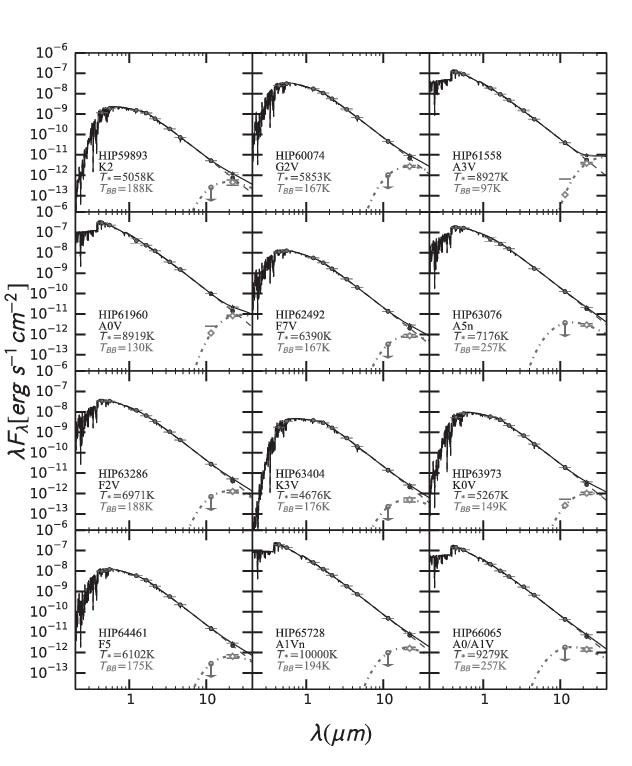}
\caption{continued.}
\end{figure*}
\clearpage

\setcounter{figure}{5}
\begin{figure*}
\centering
\includegraphics[scale=0.7]{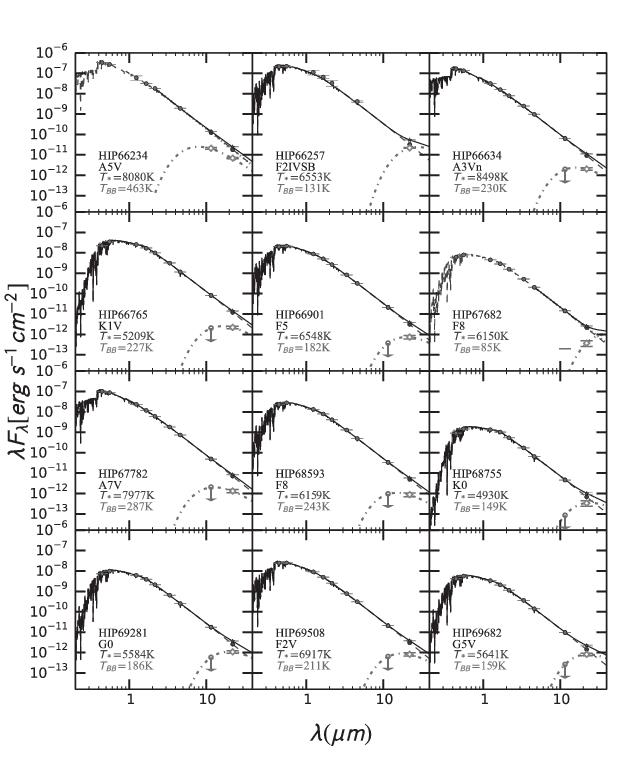}
\caption{continued.}
\end{figure*}
\clearpage

\setcounter{figure}{5}
\begin{figure*}
\centering
\includegraphics[scale=0.7]{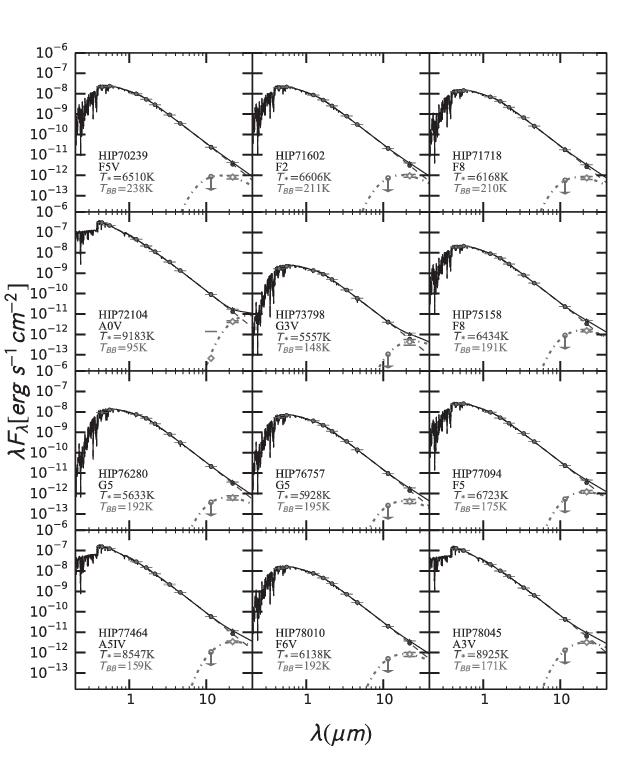}
\caption{continued.}
\end{figure*}
\clearpage

\setcounter{figure}{5}
\begin{figure*}
\centering
\includegraphics[scale=0.7]{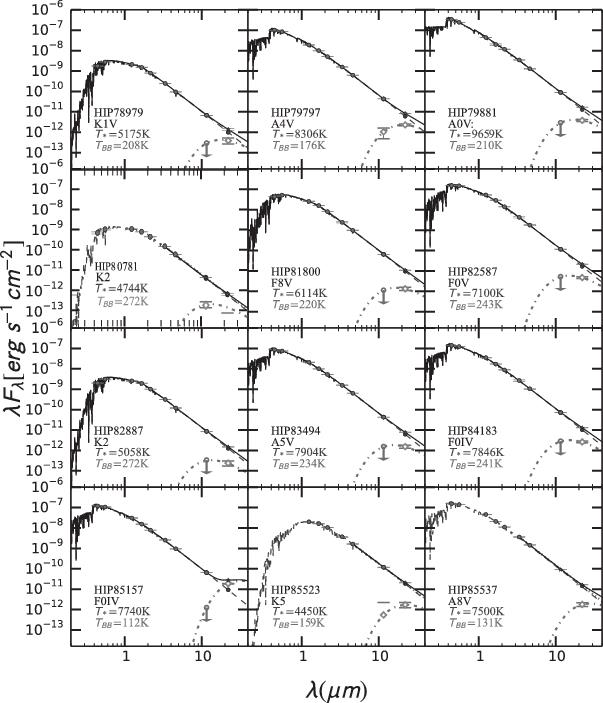}
\caption{continued.}
\end{figure*}
\clearpage

\setcounter{figure}{5}
\begin{figure*}
\centering
\includegraphics[scale=0.7]{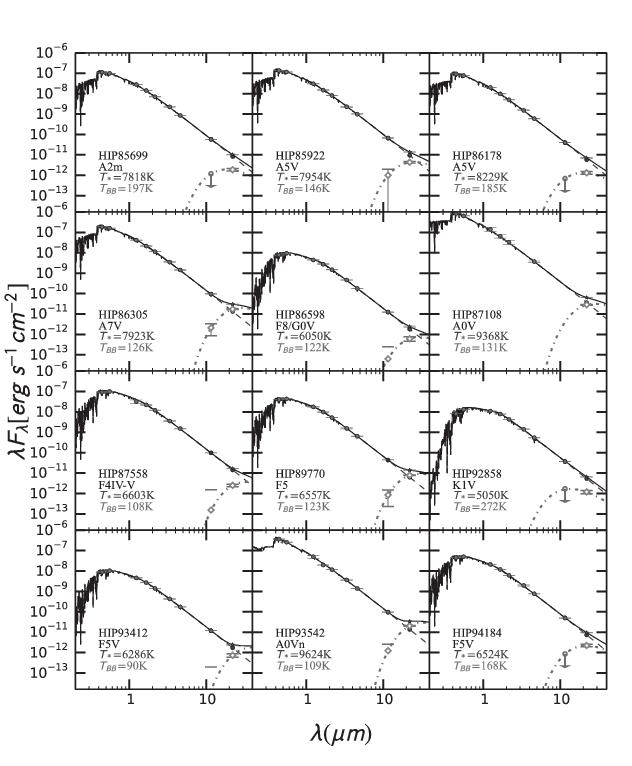}
\caption{continued.}
\end{figure*}
\clearpage

\setcounter{figure}{5}
\begin{figure*}
\centering
\includegraphics[scale=0.7]{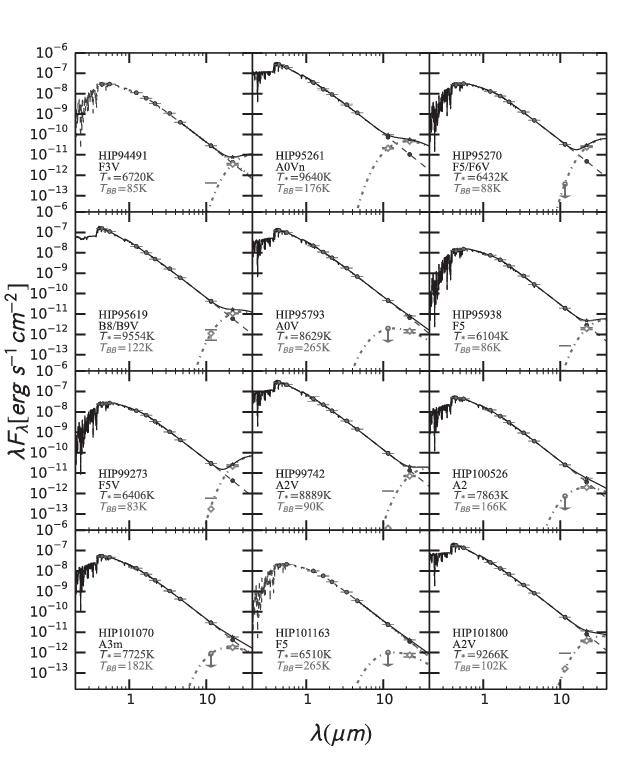}
\caption{continued.}
\end{figure*}
\clearpage

\setcounter{figure}{5}
\begin{figure*}
\centering
\includegraphics[scale=0.7]{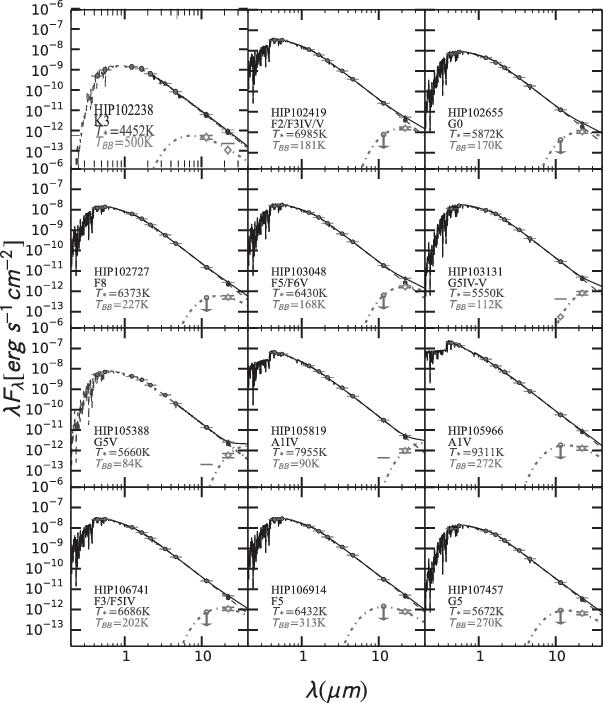}
\caption{continued.}
\end{figure*}
\clearpage

\setcounter{figure}{5}
\begin{figure*}
\centering
\includegraphics[scale=0.7]{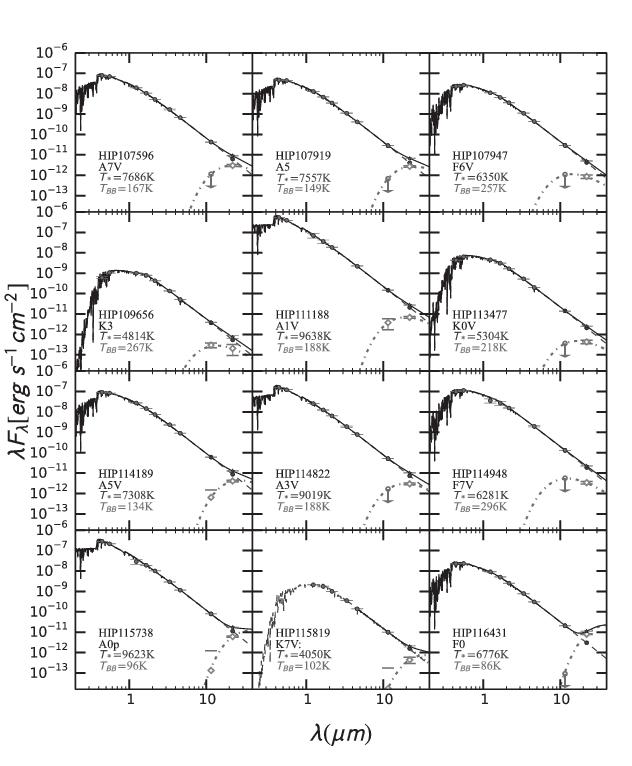}
\caption{continued.}
\end{figure*}
\clearpage

\setcounter{figure}{5}
\begin{figure*}
\centering
\includegraphics[scale=0.7]{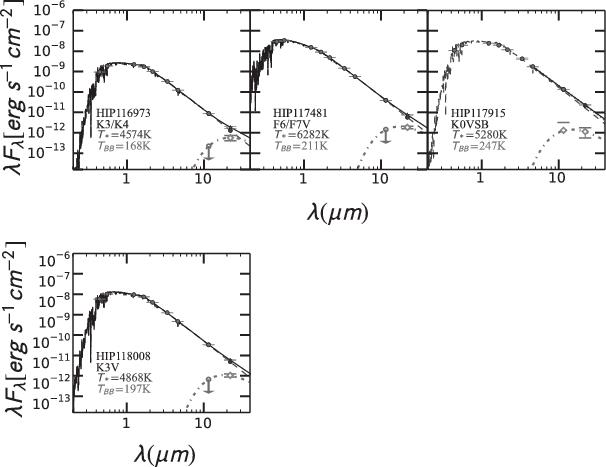}
\caption{continued.}
\end{figure*}
\clearpage

\begin{figure}
\centering
\includegraphics[scale=0.8]{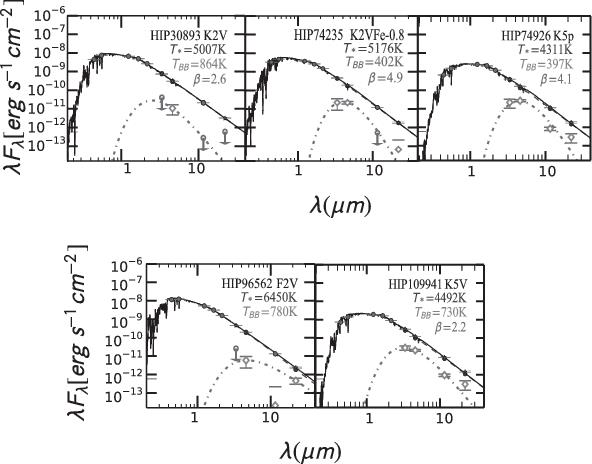}
\caption{SEDs of stars with $W2$ excesses above the 95\% confidence level. 
The stellar photosphere (dashed line) was fit to the $BVJHK_s$ photometry only. 
A blackbody was fit to the excess around the F star HIP 96562.
In the cases of the four K stars, we fit modified blackbody functions (dot-dashed lines) to the \WS\ excess fluxes (diamonds), or to the \WS\ 3$\sigma$ upper limit fluxes (open circles with downward arrows) when the excesses were negative. 
The K-star SEDs require a wide range of grain-emissivity index values ($\beta$), some of which are unphysical. The nature of these excesses remains uncertain at this time.}
\label{fig:SEDw2}
\end{figure}

\begin{figure}
\centering
\includegraphics[scale=0.6]{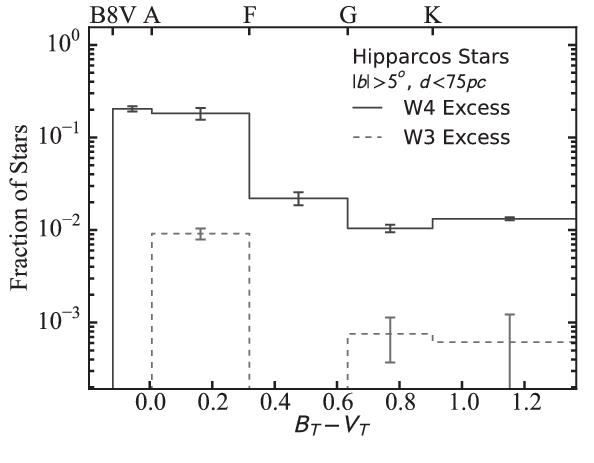}
\caption{Fraction of \WS\ excesses detected in this survey as a function of spectral type from our science sample. To determine the excess fraction at each wavelength, we chose the most sensitive color combination.}
\label{fig:wise_BV_w3w4}
\end{figure}

\begin{figure}
\centering
\includegraphics[scale=0.2]{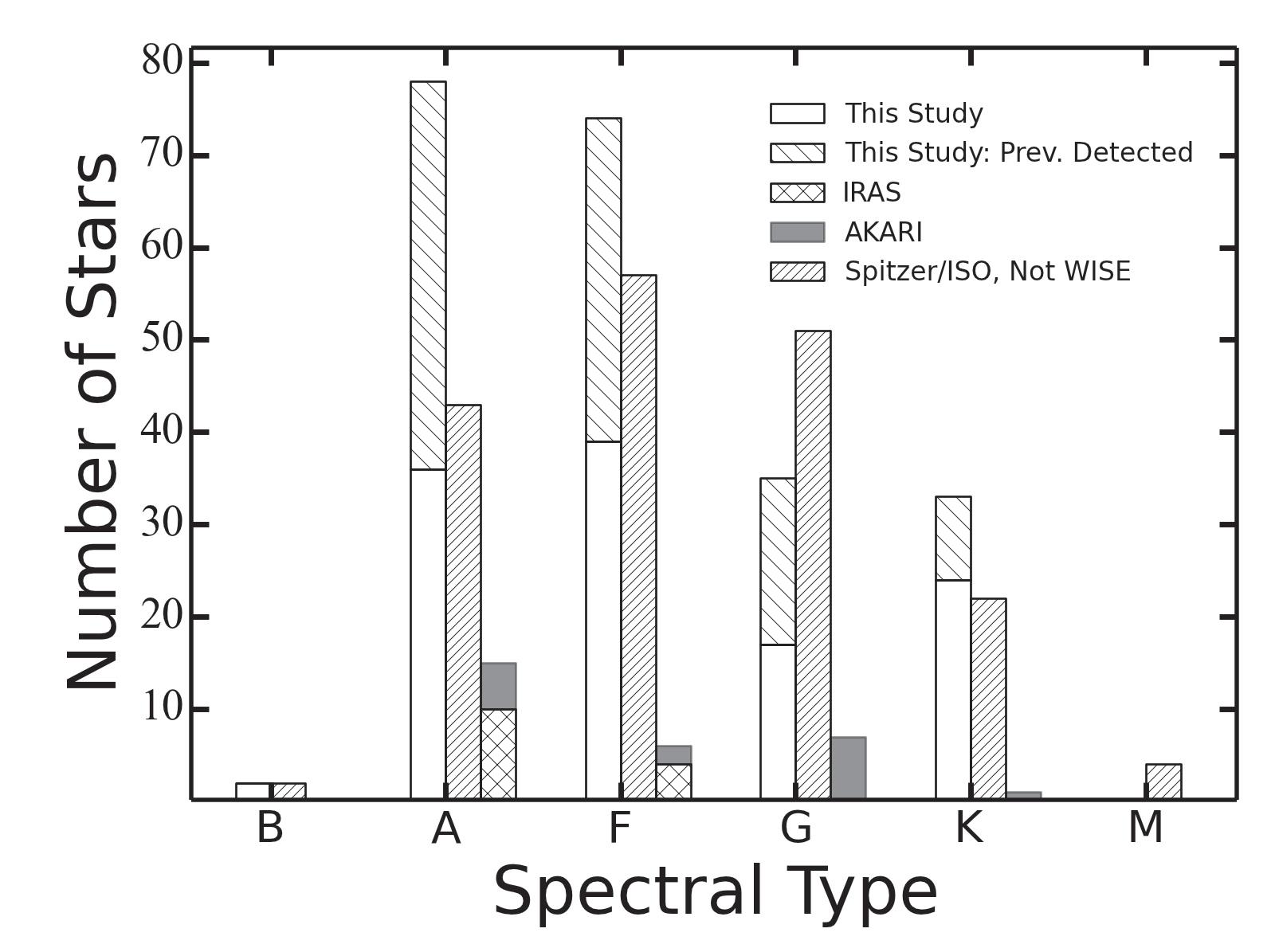}
\caption{Distribution of excesses detected as a function of spectral type using \WS\  (this paper) compared to IR excess stars detected by pointed surveys and other all-sky surveys. All the excesses are compared at wavelengths between 10--30$\micron$, for stars that are outside the galactic plane $|b|\geq 5^{\circ}$ and within 75~pc of the Sun.}
\label{fig:wise_vs_all_spt}
\end{figure}

\end{document}

%% file: table1.tex
\begin{deluxetable*}{lclc}

	\tabletypesize{\footnotesize}
	\tablewidth{0pt}
	\tablecaption{Unmatched {\it Hipparcos} Stars in the {\it WISE} All-Sky Source Catalog}
	\tablehead{\colhead{Object} & \colhead{Reason} & \colhead{Object} & \colhead{Reason}}
	\startdata
		HIP4773   &  1   &   HIP35744   &   2\\
		HIP24003  &  2   &   HIP35925   &   2\\
		HIP24188  &  2   &   HIP36051   &   2\\
		HIP26218  &  1   &   HIP36113   &   2\\ 
		HIP26220  &  3   &   HIP40215   &   2\\
		HIP26221  &  3   &   HIP46675   &   1\\
		HIP26224  &  3   &   HIP52133   &   1\\
		HIP26235  &  3   &   HIP52541   &   1\\
		HIP28868  &  2   &   HIP67207   &   1\\
		HIP29303  &  2   &   HIP73471   &   1\\
		HIP29402  &  2   &   HIP85148   &   1\\
		HIP29669  &  2   &   HIP86512   &   1\\
		HIP29761  &  2   &   HIP87022   &   1\\
		HIP30164  &  1   &   HIP88333   &   1\\
		HIP30616  &  2   &   HIP88818   &   1\\
		HIP30794  &  2   &   HIP97589   &   1\\
		HIP30964  &  2   &   HIP99001   &   1\\
		HIP31423  &  2   &   HIP107094  &   1\\
		HIP33296  &  1   &   HIP114110  &   4\\
		HIP35007  &  2   &   HIP114176  &   4\\
		HIP35681  &  2   &   HIP118182  &   1\\
	\enddata
		
	\tablecomments{Reasons: 1. Tenuous or no detections in {\it WISE} $W2$, $W3$, and $W4$ images, and a non-existent entry in the WISE All-Sky Source Catalog. 2. Partial or no {\it WISE} coverage.  3. Extensive $W3$ and $W4$ saturation because of an IR-bright surrounding nebulosity. 4. Not an astrophysical object: a {\it Hipparcos} object identified as an artifact, produced by scattered light from a nearby star.}
     \label{tab:unmatched}
\end{deluxetable*}	 

%% file: table3.tex
\begin{deluxetable*}{lcccccc}
\tablewidth{0pt}
\tabletypesize{\scriptsize}
\tablecaption{Photospheric \WS\ Colors of $-0.15<B_T-V_T<1.4$~mag Main Sequence Stars}
\tablehead{\colhead{\textbf{$B_T-V_T$}} & \colhead{$W1-W4$} & \colhead{$W2-W4$} & \colhead{$W3-W4$} & \colhead{$W1-W3$} & \colhead{$W2-W3$} & \colhead{$W1-W2$} \\ 
\colhead{(mag)} & \colhead{(mag)} & \colhead{(mag)} & \colhead{(mag)} & \colhead{(mag)}
& \colhead{(mag)} & \colhead{(mag)}} 
\startdata

\textbf{-0.16}   &  -0.070 $\pm$ 0.006   &     -0.001	$\pm$ 0.005 &  0.050	$\pm$ 0.004   &    -0.117	$\pm$ 0.004  &   -0.059	$\pm$ 0.003      &     -0.045	$\pm$ 0.004\\
\textbf{-0.14}   &  -0.070 $\pm$ 0.006   &     -0.001	$\pm$ 0.005 &  0.050	$\pm$ 0.004   &    -0.117	$\pm$ 0.004  &   -0.059	$\pm$ 0.003      &     -0.045	$\pm$ 0.004\\
\textbf{-0.12}   &  -0.070 $\pm$ 0.006   &     -0.001	$\pm$ 0.005 &  0.050	$\pm$ 0.004   &    -0.117	$\pm$ 0.004  &   -0.059	$\pm$ 0.003      &     -0.045	$\pm$ 0.004\\
\textbf{-0.10}   &  -0.065 $\pm$ 0.005   &     -0.006	$\pm$ 0.004 &  0.046	$\pm$ 0.004   &    -0.115	$\pm$ 0.003  &   -0.059	$\pm$ 0.002      &     -0.047	$\pm$ 0.003\\
\textbf{-0.08}   &  -0.056 $\pm$ 0.004   &     -0.003	$\pm$ 0.004 &  0.044	$\pm$ 0.003   &    -0.105	$\pm$ 0.002  &   -0.056	$\pm$ 0.002      &     -0.049	$\pm$ 0.002\\
\textbf{-0.06}   &  -0.054 $\pm$ 0.004   &     -0.001	$\pm$ 0.004 &  0.043	$\pm$ 0.003   &    -0.104	$\pm$ 0.002  &   -0.051	$\pm$ 0.002      &     -0.050	$\pm$ 0.002\\
\textbf{-0.04}   &  -0.043 $\pm$ 0.005   &      0.009	$\pm$ 0.004 &  0.049	$\pm$ 0.003   &    -0.091	$\pm$ 0.002  &   -0.044	$\pm$ 0.001      &     -0.044	$\pm$ 0.002\\
\textbf{-0.02}   &  -0.035 $\pm$ 0.005   &      0.011	$\pm$ 0.004 &  0.051	$\pm$ 0.003   &    -0.087	$\pm$ 0.002  &   -0.041	$\pm$ 0.001      &     -0.047	$\pm$ 0.002\\
\enddata
\tablecomments{Empirically determined \WS\ vs. $B_T-V_T$ photospheric color-color trends for all six \WS\ colors obtained from the parent sample as described in \S~\ref{sec:IRAnalysis} and shown in Figure~\ref{fig:Res}. 
The standard error of the mean for the distribution of stars in each \WS\ vs. $B_T-V_T$ bin is listed. These trends were used to correct for the photspheric color variation over the $WISE$ bands and to obtain a population of colors independent of stellar temperature.}
\label{tab:wise_bv_trends}
\end{deluxetable*}

%% file: table4.tex
\begin{deluxetable}{llc}
\tablewidth{0pt}
\tabletypesize{\footnotesize}
\tablecaption{Rejected \WS\ Excesses}

\tablehead{\colhead{HIP ID} & \colhead{WISE ID}  & \colhead{Rejection} \\ 
           \colhead{}       & \colhead{}         & \colhead{Reason}}
\startdata
HIP999    &  J001230.54+143348.0  &  2\\
HIP3121   &  J003942.53+103911.7  &  2\\
HIP3729   &  J004752.96-324520.6  &  2\\
HIP4016   &  J005129.22+563005.5  &  1\\
HIP13631  &  J025532.50+184624.2  &  1\\
HIP27114  &  J054500.36-023534.3  &  1\\
HIP40122  &  J081144.04-440200.9  &  1\\
HIP60689  &  J122617.82-512146.6  &  1\\
HIP71262  &  J143426.35-541637.8  &  1\\
HIP74045  &  J150755.93+761204.2  &  2\\
HIP76907  &  J154214.76-404922.9  &  1\\
HIP79741  &  J161628.20-364453.2  &  1\\
HIP79969  &  J161922.47-254538.9  &  1\\
HIP81181  &  J163453.29-253445.3  &  1\\
HIP82384  &  J165003.66-152534.0  &  1\\
HIP83251  &  J170055.98-314640.2  &  1\\
HIP83875  &  J170833.23-231338.7  &  1\\
HIP99542  &  J201205.89+461804.8  &  1\\

\enddata
\tablecomments{Rejection reasons:\\
    1. Contamination by nearby infrared source.\\
    2. Contamination by spectroscopic secondary component.\\}
\label{tab:rejects}
\end{deluxetable}

%% file: table5.tex
\setlength{\tabcolsep}{0.02in}
\tabletypesize{\tiny}

\begin{deluxetable*}{llccccccccccccc}

\tablewidth{0pt}

\tablecaption{Stellar Parameters of Stars with IR Excesses}

\tablehead{
\colhead{HIP} & \colhead{WISE} & \colhead{SpT\tablenotemark{a}} & \colhead{Dist.\tablenotemark{b}} & \colhead{$T_*$} & \colhead{$R_*$} & \colhead{$\chi^2_*$} & \colhead{$F_{W3}$} & \colhead{$F_{W3,*}$} & \colhead{$F_{W4}$} & \colhead{$F_{W4,*}$} & \colhead{$\Delta_{F_{W3}}/F_{W3}$\tablenotemark{d}}& \colhead{$\Delta_{F_{W4}}/F_{W4}$\tablenotemark{d}}& \colhead{$W1_{corr}$\tablenotemark{e}} & \colhead{$W2_{corr}$\tablenotemark{e}}\\ 
\colhead{ID} & \colhead{ID} & \colhead{} & \colhead{(pc)} & \colhead{(K)} & \colhead{($R_\sun$)} & \colhead{} & \colhead{(mJy)} & \colhead{(mJy)} & \colhead{(mJy)} & \colhead{(mJy)} & \colhead{}& \colhead{}& \colhead{(mag)} & \colhead{(mag)} 
} 
\startdata
544 & J000637.09+290115.4 & K0V & 14 & 5493 & 0.86 & 3.3 & 536$\pm$7 & 550 & 186$\pm$4 & 153 & -0.026 & 0.178 & 4.260$\pm$0.082 & 4.290$\pm$0.051 \\
560 & J000650.16-230627.5 & F2IV & 39 & 6789 & 1.5 & 0.38 & 234$\pm$3 & 234 & 130$\pm$3 & 64.7 & -0.001 & 0.501 & 5.220$\pm$0.072 & 5.240$\pm$0.037 \\
682 & J000825.79+063700.6 & G2V & 39 & 5845 & 1.1 & 1.2 & 106$\pm$1 & 108 & 41.6$\pm$2 & 29.9 & -0.019 & 0.281 & 6.050$\pm$0.051 & 6.100$\pm$0.022 \\
1473 & J001819.60+364706.3 & A2V & 41 & 8987 & 2 & 0.62 & 527$\pm$7 & 524 & 182$\pm$3 & 144 & 0.006 & 0.208 & 4.260$\pm$0.097 & 4.330$\pm$0.050 \\
1481 & J001826.25-632839.6 & F8/G0V & 42 & 6138 & 1.1 & 0.73 & 102$\pm$1 & 102 & 41.7$\pm$1 & 28.4 & -0.003 & 0.319 & 6.130$\pm$0.048 & 6.150$\pm$0.023 \\
1866 & J002338.01-034548.9 & K: & 47 & 4527 & 0.65 & 1.7 & 21.1$\pm$0.4 & 21.4 & 9.22$\pm$1 & 5.99 & -0.012 & 0.350 & 7.830$\pm$0.022 & 7.910$\pm$0.021 \\
2472 & J003125.12-484812.7 & A0V & 53 & 9489 & 2.1 & 1.2 & 377$\pm$6 & 374 & 130$\pm$4 & 103 & 0.010 & 0.209 & 4.760$\pm$0.075 & 4.790$\pm$0.046 \\
2710 & J003427.10-063014.9 & F2 & 41 & 6428 & 1.2 & 1.2 & 142$\pm$2 & 145 & 48.3$\pm$2 & 40.0 & -0.019 & 0.171 & 5.680$\pm$0.061 & 5.740$\pm$0.027 \\
\enddata
\tablecomments{\hip\ stars with detected mid-IR excesses at either $W2$, $W3$ and/or $W4$. Unless otherwise noted, the stellar temperature and radius were obtained from photometric fits as described in \S~\ref{sec:modeling}.  The $\chi_\ast^2$ column gives the goodness of the photospheric fit.}
\tablenotetext{a}{Spectral types for stars downloaded from \hip\ database. Stars marked with asterisks had their spectral types estimated from their $B_T-V_T$ colors using empirical color relations from \citet{Pecaut2013}.}
\tablenotetext{b}{Parallactic distances from \hip.}
\tablenotetext{c}{Stellar temperature and radius were estimated from empirical color relations from \citet{Pecaut2013} using the listed \hip\ spectral type.}
\tablenotetext{d}{The quoted fractional excesses in $W3$ and $W4$ represent the ratios of the measured excess and the estimated stellar photospheric flux in these bands.  They have not been color-corrected for the filter response, although such corrections have been applied to the estimates of the fractional bolometric luminosities $f_d$ of the dust (Section \ref{sec:modeling}, Table \ref{tab:diskcandidates}).}
\tablenotetext{e}{Saturation corrected $W1$ and $W2$ photometry (see \S~\ref{sec:wise_systematics})}
\label{tab:stellarparameters}
\end{deluxetable*}

%% file: table6.tex
\tabletypesize{\scriptsize}
\begin{deluxetable*}{lcccccccc}

\tablewidth{0pt}
\tablecaption{IR Excess Information}
\tablehead{\colhead{} &\colhead{} &\colhead{} & \multicolumn{6}{c}{Excess Significance (\ES)}\\
\cline{4-9}\\
\colhead{HIP} & \colhead{Excess} & \colhead{New?} & \colhead{$W1-W4$} & \colhead{$W2-W4$} & \colhead{$W3-W4$} & \colhead{$W1-W3$} & \colhead{$W2-W3$} & \colhead{$W1-W2$} \\ 
\colhead{ID} & \colhead{Flag} & \colhead{($12|22 \mu m$)} & \colhead{} & \colhead{} & \colhead{} & \colhead{} & \colhead{} & \colhead{} }

\startdata
544 & UNYUNU & -N & \nodata & 2.26 & 6.47 & \nodata & -1.16 & \nodata \\
560 & YYYNNN & -N & 9.22 & 15.98 & 24.30 & 0.47 & 0.77 & 0.08 \\
682 & YYYNNN & -N & 4.67 & 6.46 & 6.81 & -0.24 & -0.18 & -0.06 \\
1473 & UYYUNU & -N & \nodata & 3.29 & 6.70 & \nodata & 0.20 & \nodata \\
1481 & YYYNNN & -N & 6.97 & 9.63 & 10.17 & 0.72 & 0.57 & 0.48 \\
1866 & YYYNNN & -Y & 3.43 & 3.48 & 3.16 & 0.43 & 1.03 & -0.27 \\
2472 & YYYNNN & -Y & 3.29 & 4.64 & 4.17 & 1.46 & 2.24 & 0.12 \\
2710 & NNYNNN & -N & 1.51 & 2.96 & 3.61 & -0.78 & -0.54 & -0.43 \\
\enddata
\tablecomments{Summary of the properties of the IR excesses attributed to circumstellar excess disks at $W2, W3$ and/or $W4$ for the stars in our science sample. The \WS\  Excess Flag  indicates the combination of detections from the various colors. Each flag is a six character string that identifies whether the star has a statistically probable (Y) or insigificant (N) excess based on the order of the color analyses: $W1-W4$, $W2-W4$,$W3-W4$,$W1-W3$, $W2-W3$ and $W1-W2$. Any stars can have unlisted (U) values, indicating that the star was rejected by the selection criteria for that particular color (\S~\ref{sec:sample_definition}). `U' entires correspond to null entries in the corresponding $Wi-Wj$ \ES\ column.
Column 3 lists whether or not the star is a new detection at the $W3$ and/or $W4$ bands (12 or 22$\mu m$). The last six columns lists the significance of the excess \ES for each color.}
\label{tab:excessstats}

\end{deluxetable*}

%% file: table7.tex
\tabletypesize{\scriptsize}
\begin{deluxetable*}{lcccccccc}
\tablewidth{0pt}

\tablecaption{Debris Disk Parameters from Single-Temperature Blackbody Fits}

\tablehead{\colhead{HIP ID} & \colhead{$T_{BB}$} & \colhead{$T_{BB_{lim}}$} & \colhead{$R_{BB}$} & \colhead{$R_{BB_{lim}}$} & \colhead{$\theta$} & \colhead{$f_d$} & \colhead{$f_{d_{lim}}$} & \colhead{Notes} \\ 
\colhead{} & \colhead{(K)} & \colhead{(K)} & \colhead{(AU)} & \colhead{(AU)} & \colhead{('')} & \colhead{($10^{-5}$)} & \colhead{($10^{-5}$)} & \colhead{}}

\startdata
544 & \nodata & $<$162 & \nodata & $>$2.3 & $>$0.17 & 6.0 & $>$0.23 & b,e \\
560 & \nodata & $<$138 & \nodata & $>$8.1 & $>$0.21 & 16 & $>$0.57 & b,e \\
682 & \nodata & $<$160 & \nodata & $>$3.2 & $>$0.083 & 9.1 & $>$0.35 & b,e \\
1473 & 112 & $<$263 & 28 & $>$5.1 & 0.12--0.68 & 2.5 & $>$0.064 & c,e \\
1481 & \nodata & $<$185 & \nodata & $>$2.7 & $>$0.065 & 9.6 & $>$0.36 & b,e \\
1866 & \nodata & $<$177 & \nodata & $>$0.99 & $>$0.021 & 27 & $>$1.0 & b,e \\
2472 & 137 & $<$311 & 22 & $>$4.2 & 0.08--0.42 & 1.7 & $>$0.055 & c,e \\
2710 & \nodata & $<$208 & \nodata & $>$2.7 & $>$0.065 & 3.9 & $>$0.14 & b,e \\
\enddata
\tablecomments{A summary of the calculated disk properties of stars with $W2$, $W3$ and $W4$ excesses. Blackbody temperatures for the dust are listed \textbf{alongside} the calculated circumstellar location, projected angular extent of the dust and the fractional bolometric luminosity.\\
Notes:\\
a. $W4$-only excess: The $W3$ excess flux in this case was either saturated or $>3\sigma$ below the photosphere. A limiting temperature and radius for the dust  cannot be determined.\\
b. $W4$-only excess: The $W3$ excess flux is formally negative and an upper limit to the excess flux is used to place a $3\sigma$ limit to the dust temperature and radius.\\
c. $W4$-only excess: The $W3$ positive excess flux in this case was used to calculate a dust temperature and radius. An upper limit to the $W3$ excess flux was used to calculate a $3\sigma$ limit to the dust temperature and radius.\\
d. $W3$-only excess: The $W4$ positive excess flux in this case was used to calculate a dust temperature and radius. An upper limit to the $W4$ excess flux was used to calculate a $3\sigma$ limit to the dust temperature and radius.\\
e. Lower limit to the fractional luminosity was calculated  for a blackbody with peak emission at $\lambda=12\mu m$ as described in \S~\ref{sec:modeling}.\\
f. Lower limit to the fractional luminosity was calculated  for a blackbody with peak emission at $\lambda=22\mu m$ as described in \S~\ref{sec:modeling}.\\
g. Significant $W3$ and $W4$ excess found in these stars. Dust parameters are exact calculations\\
h. $W3$-only excess: The $W4$ excess significance in this case was undetermined as the measurement was ignored in all $W4$ analyses as its ASC measurement was $>2\sigma$ discrepant from the mean Single Frame measurement.\\
} 
\label{tab:diskcandidates}
\end{deluxetable*}

%% file: table8.tex
\tabletypesize{\footnotesize}
\begin{deluxetable*}{lllllc}
\tablewidth{0pt}

\tablecaption{Excesses Detected in Binary Systems}

\tablehead{\colhead{Star} & \colhead{Dist.} & \colhead{Binary Separation} & \colhead{Binary Separation} & \colhead{Dust Radius} & \colhead{Dust State\tablenotemark{a}} \\ 
\colhead{} & \colhead{(pc)} & \colhead{('')} &\colhead{(AU)} & \colhead{(AU)} & \colhead{} } 

\startdata
HIP544  &  42  &     189.7         &    7900  &  2.7  &  cs\\
HIP1481  &  47  &    12.6         &    590  &  36  &  cs\\
HIP2472  &  60  &    SB         &    \nodata  &  2.7  &  cb\\
HIP4016  &  44  &    40         &    1800  &  2.9  &  cs\\
HIP6679  &  29  &    132.9         &    3900  &  5.5  &  cs\\
HIP7576  &  53  &    612         &    32000  &  22  &  cs\\
HIP9141  &  61  &    0.15         &    9.2  &  3.4  &  cs\\
HIP9902  &  29  &    52.4         &    1500  &  2.8  &  cs\\
\enddata
\tablecomments{Science sample stars with debris disks in known binary systems. The binary separation was calculated using the parallactic distance and angular separations fom the Washington Double Star Catalog. Spectroscopic binaries are listed as \textbf{SB} with no known projected separation information available. }
	\tablenotetext{a}{Orbital state of the dust: ``cs" means the dust is in a cicumstellar location around the primary star, ``cb" means the dust is in a circumbinary configuration}

\label{tab:binaries}
\end{deluxetable*}